\documentclass[onecolumn,floatfix,superscriptaddress,a4paper,showpacs,showkeys,nofootinbib,preprint]{revtex4}
\usepackage[colorlinks=true,linktocpage=true,linkcolor=blue,citecolor=blue,allcolors=blue]{hyperref}
\usepackage{epsfig}
\usepackage{latexsym}
\usepackage[utf8]{inputenc}
\usepackage{xspace}
\usepackage{indentfirst}
\usepackage{enumerate}
\usepackage{color}

\usepackage{amsmath}
\usepackage{amssymb}
\usepackage[english]{babel}
\usepackage{url}

\newcommand{\muBstar}{\mu_B^*}

\newcommand{\mean}[1]{\langle #1 \rangle}

\newcommand\ddfrac[2]{\frac{\displaystyle #1}{\displaystyle #2}}

\newcommand{\eq}[1]{\begin{align} #1 \end{align}}

\begin{document}

\title{Non-congruent phase transitions
in strongly interacting matter\\
within the Quantum van der Waals model}

\author{R. V. Poberezhnyuk}
\affiliation{Bogolyubov Institute for Theoretical Physics, 03680 Kiev, Ukraine}
\affiliation{Frankfurt Institute for Advanced Studies, Giersch Science Center,
D-60438 Frankfurt am Main, Germany}
\author{V. Vovchenko}
\affiliation{Institut f\"ur Theoretische Physik,
Goethe Universit\"at Frankfurt, D-60438 Frankfurt am Main, Germany}
\affiliation{Frankfurt Institute for Advanced Studies, Giersch Science Center,
D-60438 Frankfurt am Main, Germany}
\author{M. I. Gorenstein}
\affiliation{Bogolyubov Institute for Theoretical Physics, 03680 Kiev, Ukraine}
\affiliation{Frankfurt Institute for Advanced Studies, Giersch Science Center, D-60438 Frankfurt am Main, Germany}
\author{H. Stoecker}
\affiliation{Institut f\"ur Theoretische Physik,
Goethe Universit\"at Frankfurt, D-60438 Frankfurt am Main, Germany}
\affiliation{Frankfurt Institute for Advanced Studies, Giersch Science Center,
D-60438 Frankfurt am Main, Germany}
\affiliation{GSI Helmholtzzentrum f\"ur Schwerionenforschung GmbH, D-64291 Darmstadt, Germany}

\date{\today}

\begin{abstract}
The non-congruent liquid-gas phase transition (LGPT) in asymmetric nuclear matter is studied using the recently developed Quantum van der Waals model in the grand canonical ensemble. Different values of the electric-to-baryon charge ratio, $Q/B$, are considered. 
This non-congruent LGPT exhibits several features which are not present in the congruent LGPT of symmetric nuclear matter.
These include a continuous phase transformation, a change in the location of the critical point, and the separation of the critical point and the endpoints.
The effects which are associated with the non-congruent LGPT
become negligible for the following cases: when $Q/B$ approaches its limiting values, $0.5$ or $0$, or if quantum statistical effects can be neglected.
The latter situation is realized when the particle degeneracy attains  large values, $g\gtrsim 10$.
\end{abstract}


\maketitle

\section{Introduction}
An infinite hypothetical system of interacting neutrons and protons in equal proportions is called symmetric nuclear matter.
The known phenomenology of the nucleon-nucleon interaction suggests  short range repulsion and intermediate range attraction.
This yields a first-order liquid-gas phase transition (LGPT) from a diluted (gaseous) to a dense (liquid) phase in symmetric nuclear matter, and, correspondingly, for a discontinuity of the particle number density as a function of pressure.

Experimentally, evidence for a LGPT in nuclear matter was first reported
in Refs.~\cite{Finn:1982tc,Minich:1982tb,Hirsch:1984yj}. Systematic measurements of the nuclear caloric
curve were reported by ALADIN collaboration~\cite{Pochodzalla:1995xy,LindenstruthThesis}, and other experiments~\cite{Natowitz:2002nw,Karnaukhov:2003vp}.
The thermodynamics of nuclear matter was applied to the production of nuclear fragments in heavy ion collisions in Refs. \cite{Jennings:1982wi,Ropke:1982ino,Fai:1982zk,Biro:1981es,Stoecker:1981za,Csernai:1984hf,Molitoris:1986pp,Hahn:1986mb,Aichelin:1988me,Peilert:1989kr,Peilert:1991sm}
in the 1980es (see Ref.~\cite{Csernai:1986qf} for a review of these early developments).

The present paper treats the more complex situation when the densities of neutrons and protons are not equal, i.e., the ratio of the electric-to-baryon charge, $Q/B \neq 0.5$. This asymmetric nuclear matter is the subject of Refs.~\cite{Muller:1995ji,Kolomietz:2001gd,Das:2002fr,Qian:2002kj,Sil:2004bu,Ducoin:2005aa,Iosilevskiy:2010qr,Hempel:2013tfa,Fedoseew:2014cma,Hempel:2015eoj}.

Asymmetric nuclear matter is of interest for both heavy ion-collisions and nuclear astrophysics: neutron-rich matter is present in compact stars, binary neutron star merges \cite{Hanauske:2017oxo}, and it is relevant for type-II supernova evolution~\cite{Lattimer:2000kb,Glendenning:2001pe}. 
Asymmetric nuclear matter exhibits a strong dependence on the isospin. As the $Q/B$ ratio is fixed, this additional isospin degree of freedom can not be exploited by the system in pure phases. In the mixed phase, the total asymmetry is constant, while the local asymmetries in the gaseous fraction and in the liquid fraction can be different. Indeed, it is thermodynamically favorable for the total system if the liquid fraction is more symmetric  than the gaseous one -- this isospin distillation phenomenon has been predicted \cite{Greiner:1987tg,Greiner:1988pc,Greiner:1991us} long before for the exactly analogous phenomenon dubbed strangeness distillation in the first analysis of a production mechanism for possible strangelet formation in high energy heavy ion collisions and astrophysical situations. The additional isospin degree of freedom changes the energy density. This contribution becomes negligible at end points of the mixed phase. Hence, the pressure does not stay constant, but continuously changes as the system crosses the mixed phase region, while pressures of components remain equal in every point of the mixed phase \cite{Glendenning:1992vb}. Other features of the phase transitions (PTs) are modified as well. For instance the chemical potentials show similar behavior as pressure. This leads to an additional dimension in the phase diagram.
Such PTs are called ``Gibbs PTs", or, following the more recent terminology, ``non-congruent PTs" \cite{Iosilevskiy:2010qr}.
This notion contrasts with the ``Maxwell" or ``congruent" PTs, where only one globally conserved charge is allowed.
Asymmetric nuclear matter is therefore a role model for non-congruent PTs. These are most relevant also for the conjectured PT between a hadron gas and a quark-gluon plasma. The later may occur in the course of binary neutron stars merges and in relativistic heavy-ion collisions,
where the associated  distillation process was first proposed for strangeness as a signature of that PT~\cite{Greiner:1987tg,Greiner:1988pc,Greiner:1991us}.

For asymmetric nuclear matter, the order parameter therefore is not longer given by the difference between the net baryon densities of the liquid and the gaseous phases, $n_l-n_g$, but rather by the asymmetry factor, $Q/B$ \cite{Ducoin:2005aa}. When $Q/B$ is fixed, the system moves along the PT line in the $(\mu_B, T)$-plane. This corresponds to a continuous transformation from one pure phase to the other 
\cite{Muller:1995ji}. Only in the three special cases of asymmetry, $Q/B=0,0.5$, and $1$, the phase transformation does lead to the appearance of discontinuities in thermodynamic variables, which is common for congruent PTs.

The properties of nuclear matter can be described by a variety of different models.
Here we employ an extension of the classical van der Waals (vdW) model which was recently generalized to include the effects of quantum statistics, special relativity, grand canonical ensemble, and mixtures of different sized constituents. This Quantum vdW (QvdW) model has been further developed 
and applied to the description of symmetric nuclear matter in Refs.~\cite{Vovchenko:2015vxa,Vovchenko:2015xja,Vovchenko:2015pya,Redlich:2016dpb,VovchenkoThesis,Vovchenko:2017cbu,Vovchenko:2017ayq}.
The QvdW approach models the repulsive interactions by the excluded-volume corrections, while the attractive interactions are modeled by a density-proportional mean field.
The QvdW model describes the basic properties of nuclear matter rather well, the results are similar to the Walecka model~\cite{Poberezhnyuk:2017yhx}.
A generalized QvdW-type formalism, based on models of real gases equations of state 
allows variations of the excluded-volume effects and of the attractive mean field~\cite{Vovchenko:2017cbu}.
The multi-component QvdW formalism, employed in the present work, allows for the study of systems with arbitrary numbers of different components~\cite{Vovchenko:2016rkn,Vovchenko:2017zpj}. 

The present paper studies the LGPT in asymmetric nuclear matter using the QvdW model.
Section~\ref{sec:formulation} introduces the QvdW model with separate baryonic and electric chemical potentials for neutrons and protons.  Section~\ref{sec:congruent} considers four special cases with congruent LGPTs, namely, the limiting cases of: 1) symmetric nuclear matter, $Q/B=0.5$; 2) the extreme asymmetry, $Q/B=0$; 3) an arbitrary $Q/B$ ratio in the Boltzmann approximation; 4) an arbitrary $Q/B$ ratio and infinite degeneracy.
Section~\ref{sec:noncongruent} studies the general case of a non-congruent LGPT in nuclear matter with intermediate values of the asymmetry factor, $0<Q/B<0.5$, and with physical values of (spin) degeneracy. 
Section~\ref{sec:fluct} presents the calculation of the susceptibilities of the fluctuations of the baryonic and of the electric charges for asymmetric nuclear matter. 
Section~\ref{sec:isospin-dep} considers the non-congruent LGPTs within QvdW model generalized to take into account the isospin-dependencies of the attractive and repulsive parameters.
A summary in Sec.~\ref{summary} closes the article.

\section{Nuclear matter with two different conserved charges}
\label{sec:formulation}

We consider an infinite system of interacting nucleons consisting of neutrons and protons which differ only by the electric charge they carry. The total baryonic, $B$, and electric, $Q$, charges of the system in the grand canonical ensemble are regulated by the corresponding chemical potentials, $\mu_B$ and $\mu_Q$. Then $\mu_n=\mu_B$ is the chemical potential of the neutrons and $\mu_p=\mu_B+\mu_Q$ is the chemical potential of the protons. The QvdW model yields the total pressure of the nucleons as~\cite{Vovchenko:2017ayq}:
\begin{eqnarray}
 p(T,\mu_B,\mu_Q) ~=~ p_{n}^{\rm id}( T,\muBstar)~+~ p_{p}^{\rm id}( T,\muBstar + \mu_Q )
~-~a~n_B^{2}
~. \label{pB}
\end{eqnarray}%

Here $T$ is the temperature, $p_{n}^{\rm id}$, $p_{p}^{\rm id}$ are the pressures of the ideal Fermi gas of the neutrons and the protons, respectively. $\mu_B^*$ is the shifted baryon chemical potential due to the QvdW interactions:
\eq{\label{muB*}
\muBstar = \mu_B - b\,p - a\,b\,n_B^2 + 2\,a\,n_B~.~~~~~~
}
Here it is assumed that the repulsive excluded volume terms and the mean field attraction terms of protons and neutrons do {\it not} differ, and that their masses do {\it not} differ either.
The interaction parameters, $a$ and $b$, yield, respectively, the strength of the attraction and of the repulsion between the nucleons. 
As the interactions between all protons and neutrons  here are assumed to be the same,
also the shift in the chemical potential is the same for both, protons and neutrons.
The densities of the baryonic and the electric charges are given by the partial derivatives of the pressure with respect to the corresponding chemical potentials,
\eq{\label{nB}
n_B(T,\mu_B,\mu_Q)~=~\left[\frac{\partial p}{\partial \mu_B}\right]_{T,\mu_Q}~=~\frac{n_{n}^{\rm id}(T,\muBstar)~+~n_{p}^{\rm id}(T,\muBstar + \mu_Q)}{1~+~b~[n_{n}^{\rm id}(T,\muBstar)~+~n_{p}^{\rm id}(T,\muBstar + \mu_Q)]}~,\\\label{nQ}
n_Q(T,\mu_B,\mu_Q)~=~\left[\frac{\partial p}{\partial \mu_Q}\right]_{T,\mu_B}~=~\frac{n_{p}^{\rm id}(T,\muBstar + \mu_Q)}{1~+~b~[n_{n}^{\rm id}(T,\muBstar)~+~n_{p}^{\rm id}(T,\muBstar + \mu_Q)]}~.
}
Here $n_{n}^{\rm id}=n_{n}^{\rm id}(T,\mu _{B}^*)$ and $n_{p}^{\rm id}=n_{p}^{\rm id}(T,\mu _{B}^*+\mu_Q)$ are the ideal gas densities of neutrons and protons, respectively.
The pressure and the density of the ideal Fermi gas of neutrons, $j=n$, and protons, $j=p$, are given by
\eq{
p_j^{\rm id}(T,\mu^*_j)~ &=~
 \frac{ g_j}{6\pi^2} \int_0^{\infty} k^2\,dk
\frac{ k^2}{\sqrt{m_j^{2} + k^2}}\, f_{\rm k}(T,\mu^*_j)  \,,
\label{p-id}\\
n_j^{\rm id}(T,\mu^*_j)~ &=~
 \frac{ g_j}{2\pi^2} \int_0^{\infty} k^2\,dk
\, f_{\rm k}(T,\mu^*_j)  \,.
\label{n-id}
}
The density of states corresponding to the momentum k is given by
\eq{
 f_{\rm k}(T,\mu^*_j)\, = \, 
\left[\exp{ \left( \frac{\sqrt{ m_j^{2}+k^2} - \mu^*_j}{T}\right)} + 1\right]^{-1}\,.
\label{quantum-f}
}
$g_j$ is the number of the internal quantum states -- the degeneracy factor of the neutrons and the protons, which are spin 1/2 particles, therefore $g_n=g_p=2$. 
The masses of both, neutrons and protons, are assumed to be equal, with
$m_n = m_p = 938$~MeV.

As the values of the charges, $B$ and $Q$, are conserved, the total system is also required to have the charge ratio $Q/B$,
\eq{\label{conserv-general}
\frac{n_{Q}}{n_B}~=~\frac{n_p^{\rm id}(T,\muBstar + \mu_Q)}{n_n^{\rm id}(T,\muBstar)~+~n_p^{\rm id}(T,\muBstar + \mu_Q)}~=~\frac{Q}{B}~=~\text{const}~.
}

The thermodynamical functions at given $T$- and $\mu_B$-values are calculated by solving the self-consistent system of the four transcendental equations, (\ref{pB}),(\ref{muB*}),(\ref{nB}), and (\ref{conserv-general}), with respect to the four unknown quantities, $\mu^*$, $\mu_Q$, $p$, and $n_B$.



\section{Four special cases for congruent phase transitions}
\label{sec:congruent}

\subsection{Symmetric nuclear matter (g~=~4)}
\label{sec:congruent-A}

The LGPT in symmetric nuclear matter  was studied  within the QvdW model in Ref.~\cite{Vovchenko:2015vxa}.
The case of symmetric nuclear matter
corresponds to a fixed value of the baryon-to-charge ratio, $Q/B=0.5$. In this case, as follows from Eqs.~(\ref{pB}),(\ref{nB}), and (\ref{conserv-general}), the electric chemical potential is always zero, $\mu_Q\equiv0$, and the multiplicities of both neutrons and protons are regulated by a single chemical potential, $\mu_B$. Hence, the system of nucleons is a single-component system. In this case, the degeneracy factor of the nucleons is to be taken as $g=4$, which includes two (isospin) charge states and two spin states.

The QvdW interactions are taken to be equal for all pairs of nucleons: $\ a=329$ MeV fm$^{3}$ for the attractive term and $b=3.42$ fm$^{3}$ for the repulsive term.
These $a$ and $b$ values
were obtained in Ref.~\cite{Vovchenko:2015vxa} by fitting the
binding energy
and the saturation density
of the ground state (GS; $T=0$, $p=0$) of symmetric nuclear matter:
\eq{\label{propertiest0}
 \varepsilon^{\rm GS} / n^{\rm GS}_B ~\cong~ m + E^{\rm GS}_b ~\cong~ 922~{\rm MeV}~,~~~~~~~~~~~
 n^{\rm GS}_B ~\cong ~ 0.16~{\rm fm}^{-3}~.
}

The position of the critical point (CP) of symmetric nuclear matter within QvdW model is \cite{Vovchenko:2015vxa}: 
\eq{
T_c~=~19.7~{\rm MeV}~,~~~~~n_c~=~0.072~{\rm fm^{-3}}~,~~~~~p_c~=~0.52~{\rm MeV~fm^{-3}}~.
}

\begin{figure}[h!]
\includegraphics[width=0.49\textwidth]{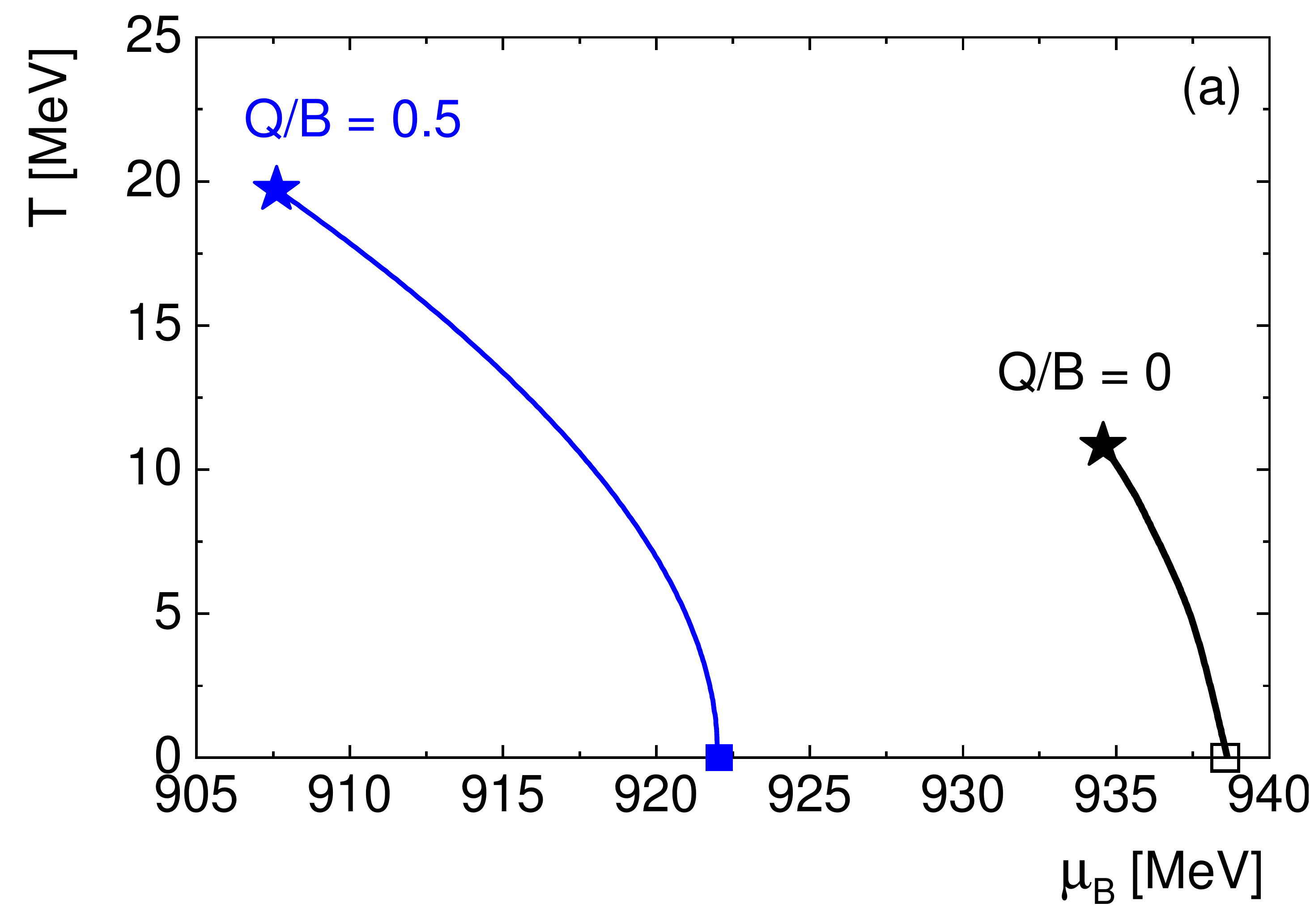}
\includegraphics[width=0.49\textwidth]{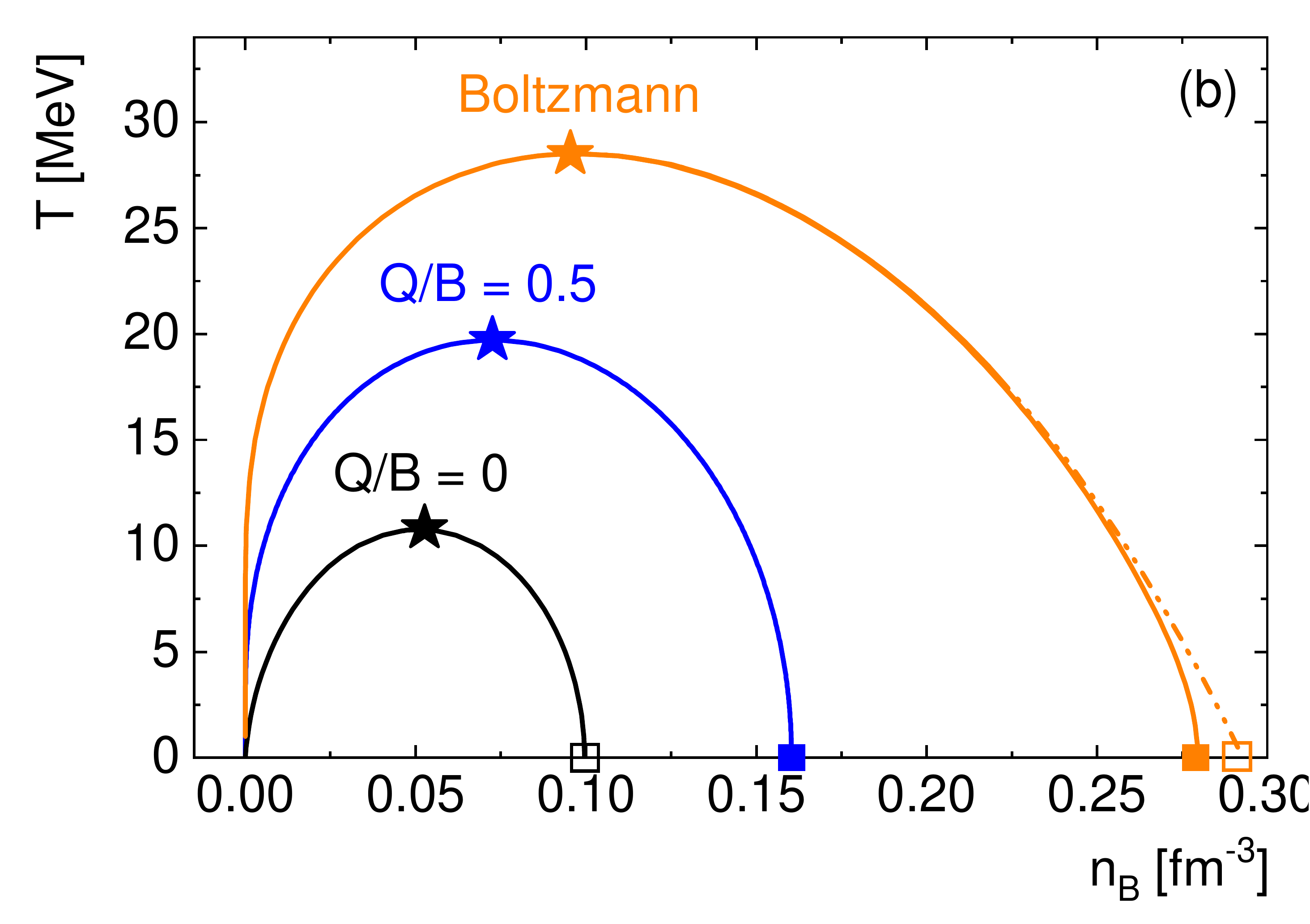}
\caption{\label{congr} ($a$) Congruent LGPT lines in the ($\mu_B$, $T$) coordinates  and ($b$) congruent LGPT regions in the ($n_B$, $T$) coordinates  within the QvdW model for nucleons with  degeneracy  factors $g=4$ (symmetric nuclear matter; blue curves) and $g=2$ (pure neutron matter; black curves). The orange solid and dash-dotted curves in ($b$) correspond to a huge degeneracy factor $g=1000$ and to the Boltzmann approximation (with arbitrary $g$ or $Q/B$), respectively. Critical points and ground states are represented, respectively, by the stars and the squares.
}
\end{figure}

The LGPT line in the ($\mu_B$, $T$) coordinates as well as the LGPT region in the ($n_B$, $T$) coordinates were obtained in Ref.~\cite{Vovchenko:2015vxa} within the QvdW model for symmetric nuclear matter. They are presented in Fig.~\ref{congr} $(a)$ and $(b)$, respectively, by the solid blue curves. The CP and the GS are represented by the blue star and the blue square, respectively. 

\subsection{Neutron matter (g~=~2)}
Another limiting case is the completely asymmetric nuclear matter  consisting of neutrons only \footnote{Note that neutron matter as discussed here does not match neutron star matter -- the latter must include light and heavy nuclei, beta-equilibrium, leptons, strange hadrons, and eventually also a quark matter contribution. So the term neutron matter is used here only for a hypothetical state of neutrons only.}. 
This corresponds to a zero asymmetry parameter, $Q/B=0$, implying $\mu_Q \to -\infty$.
In analogy to the symmetric nuclear matter case, the system is described by the single-component QvdW equation, but with a smaller degeneracy factor, $g = 2$, which counts only the two spin states of the neutron.
The same values of the interaction parameters $a$ and $b$ are used as in the case of the nucleon-nucleon interaction in symmetric nuclear matter. 

The values of the thermodynamic quantities in the nuclear GS of symmetric nuclear matter are fixed to known empirical values, see Eq.~(\ref{propertiest0}).
These values fix parameters $a$ and $b$ for nucleons, as well as the value of the shifted chemical potential in the GS, $\mu^*_{\rm GS} =  998~{\rm MeV}$ ($\mu_{\rm GS} =  921.5~{\rm MeV}$)~\cite{Vovchenko:2015vxa}.
As the degeneracy factor for pure neutron matter is twice smaller than that of symmetric nuclear matter,
 different GS properties for pure neutron matter are expected as compared to the symmetric nuclear matter.
A straightforward calculation yields: $\mu^*_{\rm GS} =  993~{\rm MeV}$ ($\mu_{\rm GS} =  938.45~{\rm MeV}$) and
\eq{\label{propertiest0-g2}
 \varepsilon^{\rm GS} / n^{\rm GS}_B ~\cong~ m + E^{\rm GS}_b ~\cong~ 938.33~{\rm MeV}~,~~~~~~~~~~~
 n^{\rm GS}_B~\cong ~ 0.10~{\rm fm}^{-3}~.
}

One sees that the neutrons binding energy in the GS is positive, $E^{\rm GS}_b=0.33~{\rm MeV}>0$, thus the GS of pure neutron matter does exist in this model but neutron matter is not self-bound.
The position of the CP is determined by the following equations,
\eq{\label{cp}
\left(\frac{\partial p}{\partial n_B}\right)_T~=~0,~~~~~~~~~~~~~~\left(\frac{\partial^2 p}{\partial n_B^2}\right)_T~=~0~,
}
which give,
\eq{
T_c~=~10.8~{\rm MeV}~,~~~~~n_c~=~0.051~{\rm fm^{-3}}~,~~~~~p_c~=~0.197~{\rm MeV~fm^{-3}}~.
}

Mixed phase boundaries at given $T<T_c$ are derived from the Gibbs equilibrium condition:
\eq{\label{gibbs-congr}
p_G(T,\mu_B)~=~p_L(T,\mu_B)~,
}
where $p_G(T,\mu_B)$ and $p_L(T,\mu_B)$ are two solutions for pressure with different shifted chemical potentials, $\mu^{*G}_{B}\neq\mu^{*L}_{B}$. These two solutions correspond to pressures on the boundaries of pure gaseous and liquid phases, respectively.

The LGPT lines in the ($\mu_B$, $T$)-coordinates and the LGPT regions in the ($n_B$, $T$)-coordinates are presented for neutron matter in Fig.~\ref{congr} $(a)$ and $(b)$, respectively, solid black curves.  The CP is depicted by the black star while the not-self bound GS is represented by the open black squares.
In section \ref{sec:isospin-dep} it is shown that the consideration of isospin-dependent parameters within QvdW model results in absence of both GS and LGPT in pure neutron matter.

\subsection{Boltzmann approximation (arbitrary g)}
In the Boltzmann approximation, Eqs.~(\ref{pB})-(\ref{nB}) are reduced to the classical vdW equation of state \cite{Vovchenko:2015vxa}:
\eq{
p_B(T,n_B)~=~\frac{n_B~T}{1~-~b~n_B}~-~a~n_B^2~,
}
which is independent of $g$-, $m$-, and $Q/B$-values \cite{Vovchenko:2015xja}. 
The position of the CP is determined solely by the interaction parameters $a$ and $b$ \cite{landau2013statistical}:
\eq{\nonumber
&T_c^{\rm boltz}~=~\frac{8a}{27b}~=~28.54~{\rm MeV}~,~~~~~n_c^{\rm boltz}~=~\frac{1}{3b}~=~0.098~{\rm fm^{-3}}~,\\\label{cp-boltz}
&p_c^{\rm boltz}~=~\frac{a}{27b^2}~=~1.045~{\rm MeV~fm^{-3}}~.
}

Hence, the non-congruence of the LGPT vanishes\footnote{Note that this is true only if the isospin-independent interaction parameters are considered as in the present section. Within a more general QvdW formalism, which is described in section \ref{sec:isospin-dep}, the non-congruent PT takes place even in the Boltzmann approximation.}
for arbitrary $Q/B$ and $g$ values, and the position of the CP is given by 
Eq.~(\ref{cp-boltz}). 
Figure~\ref{congr} $(b)$ shows the LGPT region in the ($n_B$,~$T$) coordinates for nuclear matter with the classical vdW equation of state (dashed-dotted orange lines). Both the $Q/B$- and $g$-values given here are arbitrary. 
Note that the Boltzmann approximation is not valid at low temperatures. For instance, the entropy density in the classical vdW model becomes negative at sufficiently low temperatures, $T<T_{min}$~\cite{Vovchenko:2015xja}. 
Therefore, nuclear matter at temperatures, $T<T_{min}$, including the GS, cannot be described in the Boltzmann approximation.

\subsection{Large number of internal states (g$>>$1)}

In Sec.~III.B it was shown that a decrease in the number of the internal degrees of freedom leads to a decrease of the $T_c,~ n_c$, and $p_c$ values. This demonstrates the increased importance of the Fermi statistics.
Correspondingly, increasing $g$ will reduce the importance of the quantum statistics.
From Eq.~(\ref{nB}) it follows that, for a constant value of $n_B$ and $T$, the chemical potential must decrease when $g$ increases. If $g\rightarrow \infty$ then $\mu_B\rightarrow -\infty$ and quantum statistics can be neglected.
Hence, the non-congruence of a LGPT vanishes at $g\rightarrow \infty$ for arbitrary fixed $Q/B$. The phase diagram becomes indistinguishable from  phase diagram of the the corresponding classical vdW in the ($n_B$, $T$) coordinates. Namely,
\eq{
T_c,~ n_c,~ p_c ~\rightarrow~ T^{\rm Boltz}_c,~ n^{\rm Boltz}_c,~ p^{\rm Boltz}_c~~~~~{\rm at}~~~~~ g ~\rightarrow~ \infty~,
}
where the critical values for the Boltzmann case are given by 
Eq.~(\ref{cp-boltz}).

A QvdW system of nucleons
with a large number of internal degrees of freedom, $g=1000$, 
exhibits a CP at
\eq{
T_c~=~28.50~{\rm MeV}~,~~~~~n_c~=~0.098~{\rm fm^{-3}}~,~~~~~p_c~=~1.042~{\rm MeV~fm^{-3}}~,
}
close to the position of the classical vdW CP (\ref{cp-boltz}).

The nuclear GS is at $\mu^*_{\rm GS} =  949.5~{\rm MeV}$ ($\mu_{\rm GS} =  852.4~{\rm MeV}$), with
\eq{\label{propertiest0-g2}
 \varepsilon^{\rm GS} / n^{\rm GS}_B ~\cong~ m + E^{\rm GS}_b ~\cong~ 852.4~{\rm MeV}~,~~~~~~~~~~~
 n^{\rm GS}_B~\cong ~ 0.28~{\rm fm}^{-3}~.
}
The value of saturation density, $n^{\rm GS}_B \cong  0.28~{\rm fm}^{-3}$, is close to the corresponding value for the Boltzmann approximation, $n^{\rm GS}_B \cong  0.29~{\rm fm}^{-3}$. 
However, in the quantum statistics case, the entropy density in the GS is positive and, therefore, the GS is thermodynamically consistent, in contrast to the classical case.

The LGPT regions in ($n_B,T$) coordinates for baryons with the QvdW equation of state and the degeneracy factor $g=1000$ are shown in Fig.~\ref{congr} $(b)$ by the solid orange lines. The GS is noted by a full orange square.

The position of the CP in the QvdW model with large degeneracy is close to the position of the CP in the Boltzmann approximation even for smaller values of $g$, e.g., $T_c=27.76$~MeV for $g=10$. Hence, systems of particles with large numbers of internal states are insensitive to the effects of quantum statistics. The picture of the PT in coordinates of $T$, $n_B$, or $p$ has no dependence on $g$. An example of the system with $g>10$ are the $\Delta(1232)$ baryons from the SU(3) decuplet. 
Four isospin states, each with four spin states, yield a total degeneracy of $g_{\Delta} = 16$ for the spin-isospin degenerate symmetric matter of $\Delta(1232)$ baryons.

\section{Non-congruent phase transition}
\label{sec:noncongruent}

\begin{figure}[h!]
\includegraphics[width=0.49\textwidth]{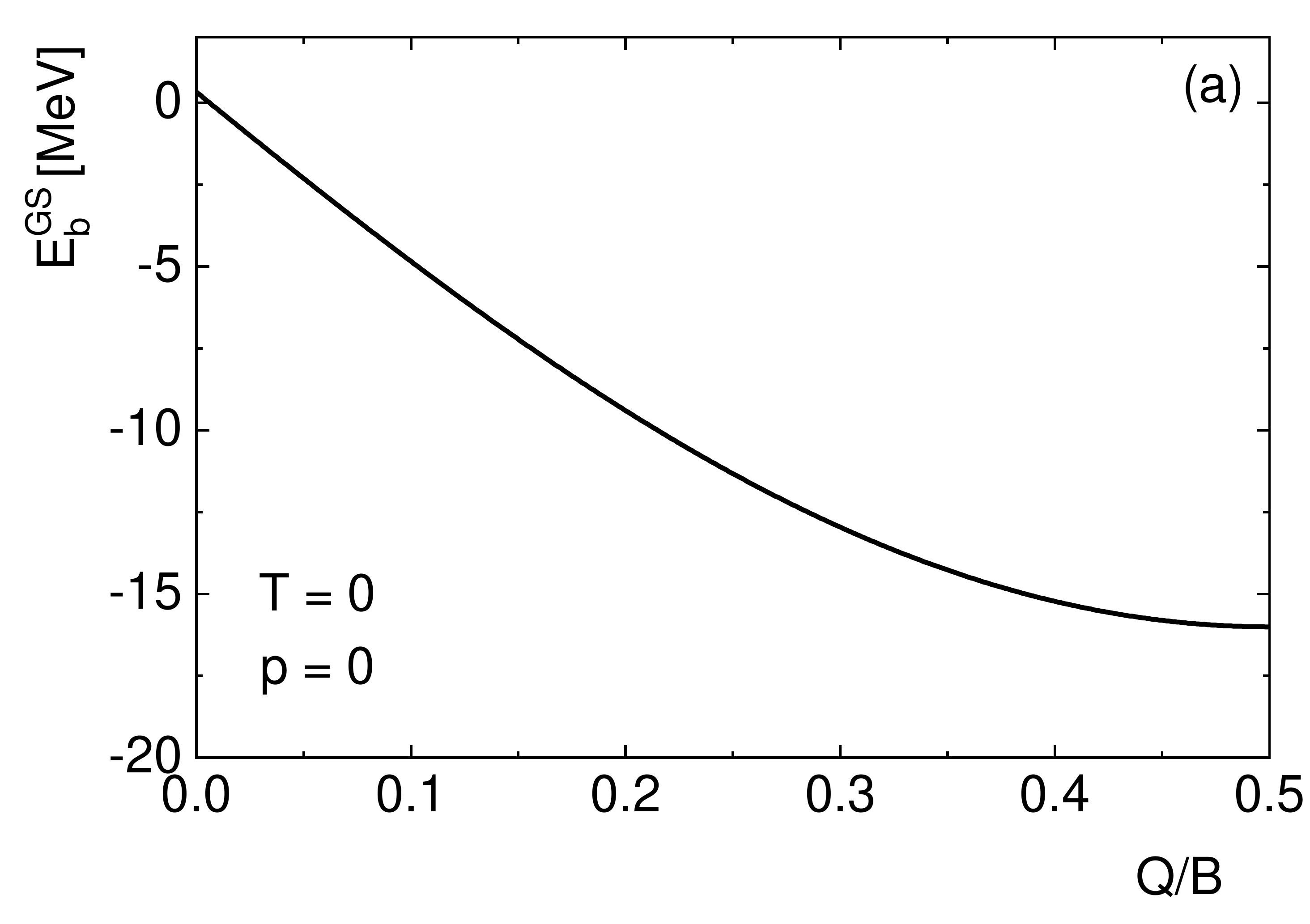}
\includegraphics[width=0.49\textwidth]{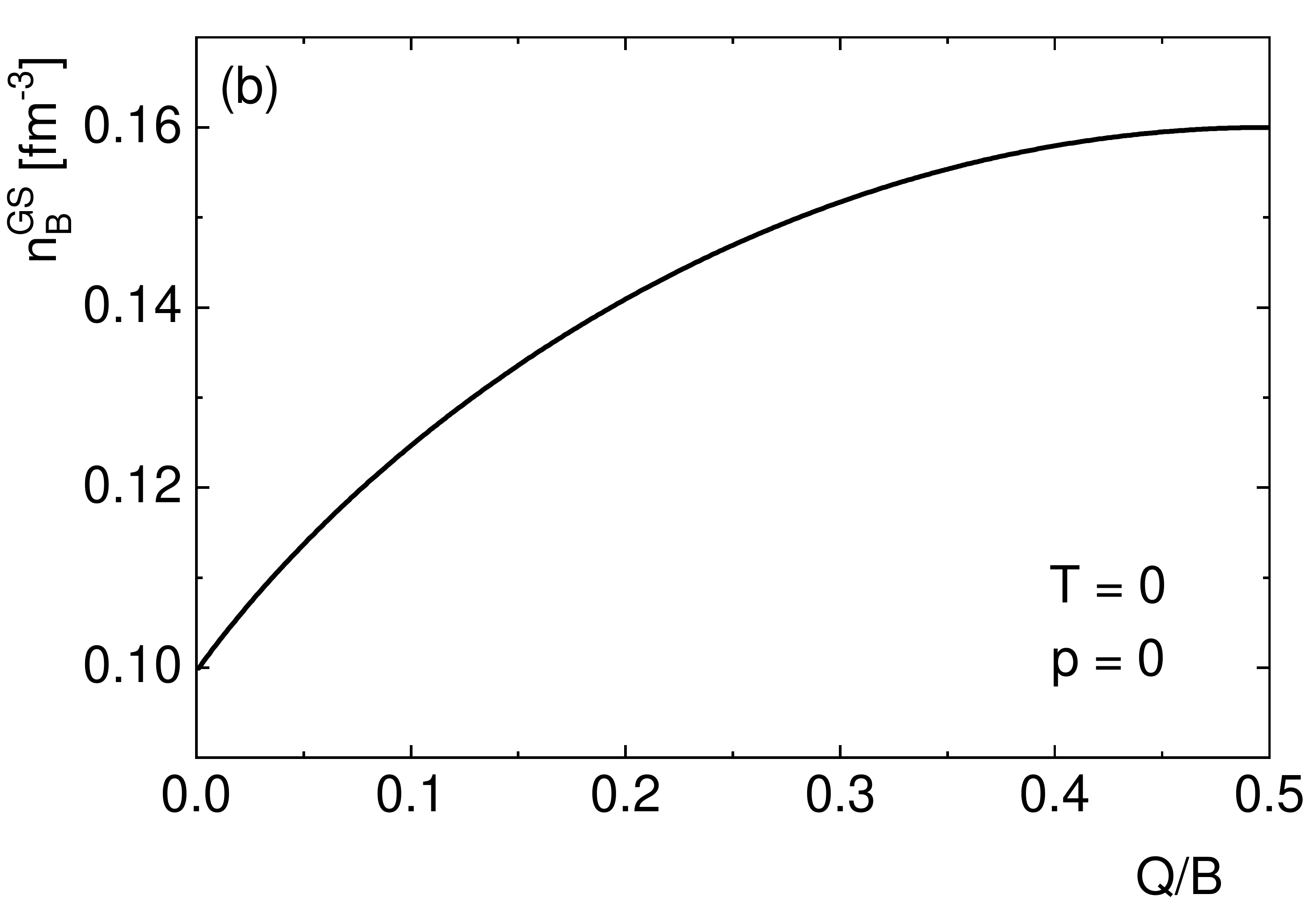}
\caption{\label{eb} Dependence on the asymmetry parameter of ($a$) the ground state binding energy 
and ($b$) of the baryon density.
}
\end{figure}

\begin{figure}[h!]
\includegraphics[width=0.55\textwidth]{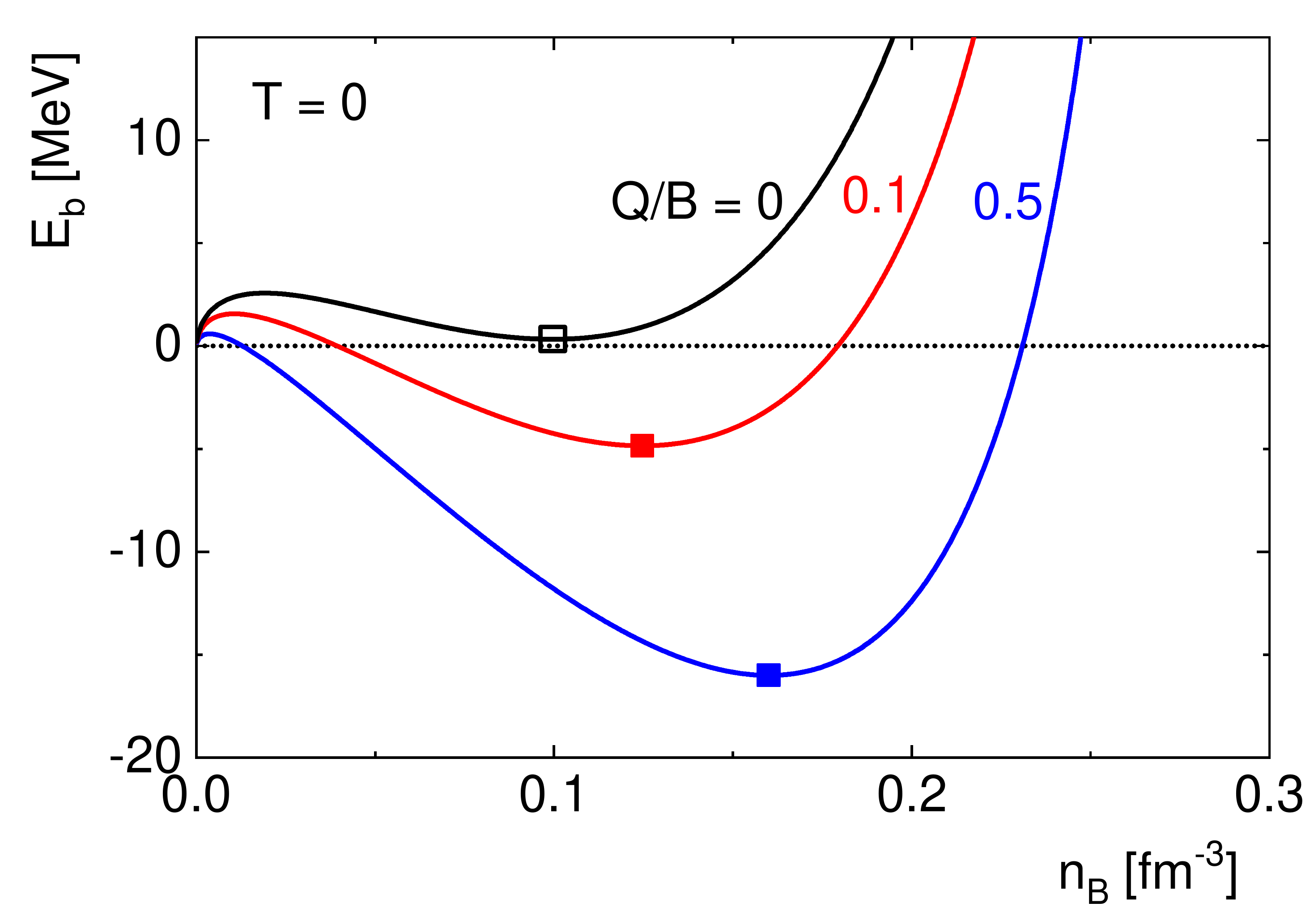}
\caption{\label{eb-n} Dependence of the binding energy on baryon density at zero temperature for the constant values of the asymmetry parameter. The ground states are represented by the squares.
}
\end{figure}

The so-called non-congruent LGPTs occur for $0<Q/B<0.5$. In this case $\mu_Q$ is finite, and two conserved charges $B$ and $Q$ must be considered, which are tuned by the corresponding chemical potentials, $\mu_B$ and $\mu_Q$.
The electric-to-baryon charge ratio is kept fixed, $Q/B= {\rm const}$.
The location of the GS is defined by two equations: $p(T=0,\mu^{\rm GS}_{B},\mu^{\rm GS}_{Q})=0$ and $Q/B={\rm const}$. 
The GS binding energy, $E^{\rm GS}_b$, and the baryon density, $n_B^{\rm GS}$, are shown in Fig.~\ref{eb} $(a)$ and $(b)$, respectively, as functions of $Q/B$. For $Q/B<0.006$, the binding energy in the GS is positive, therefore, the GS is not self-bound. 
The special case of pure neutron matter, $Q/B=0$, without beta equilibrium and without leptons, is considered in Sec.~II.B. Figure~\ref{eb-n} shows the dependence of the binding energy on $n_B$ at zero temperature for three constant values of the asymmetry parameter.
\begin{figure}[h!]
\includegraphics[width=0.49\textwidth]{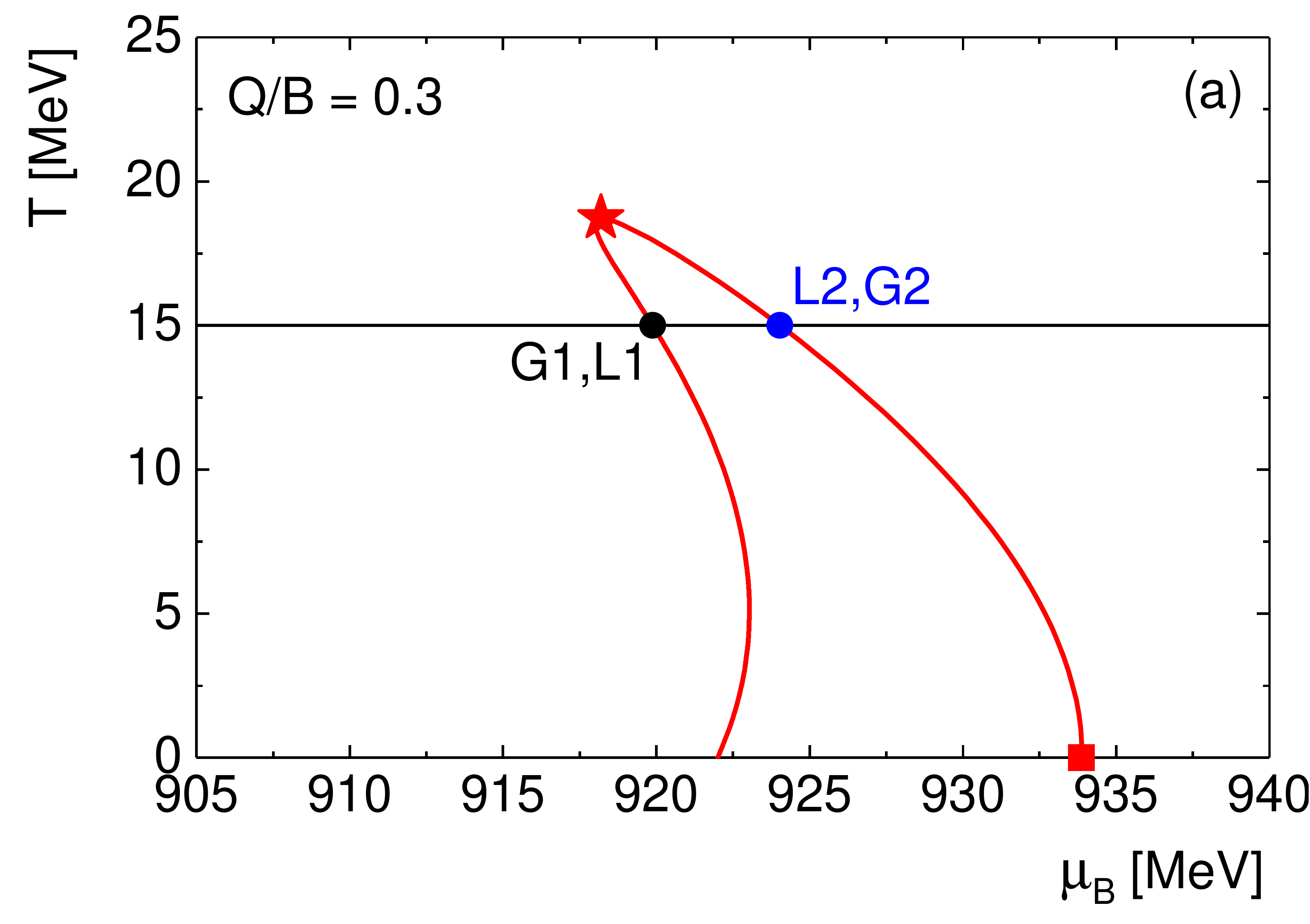}
\includegraphics[width=0.49\textwidth]{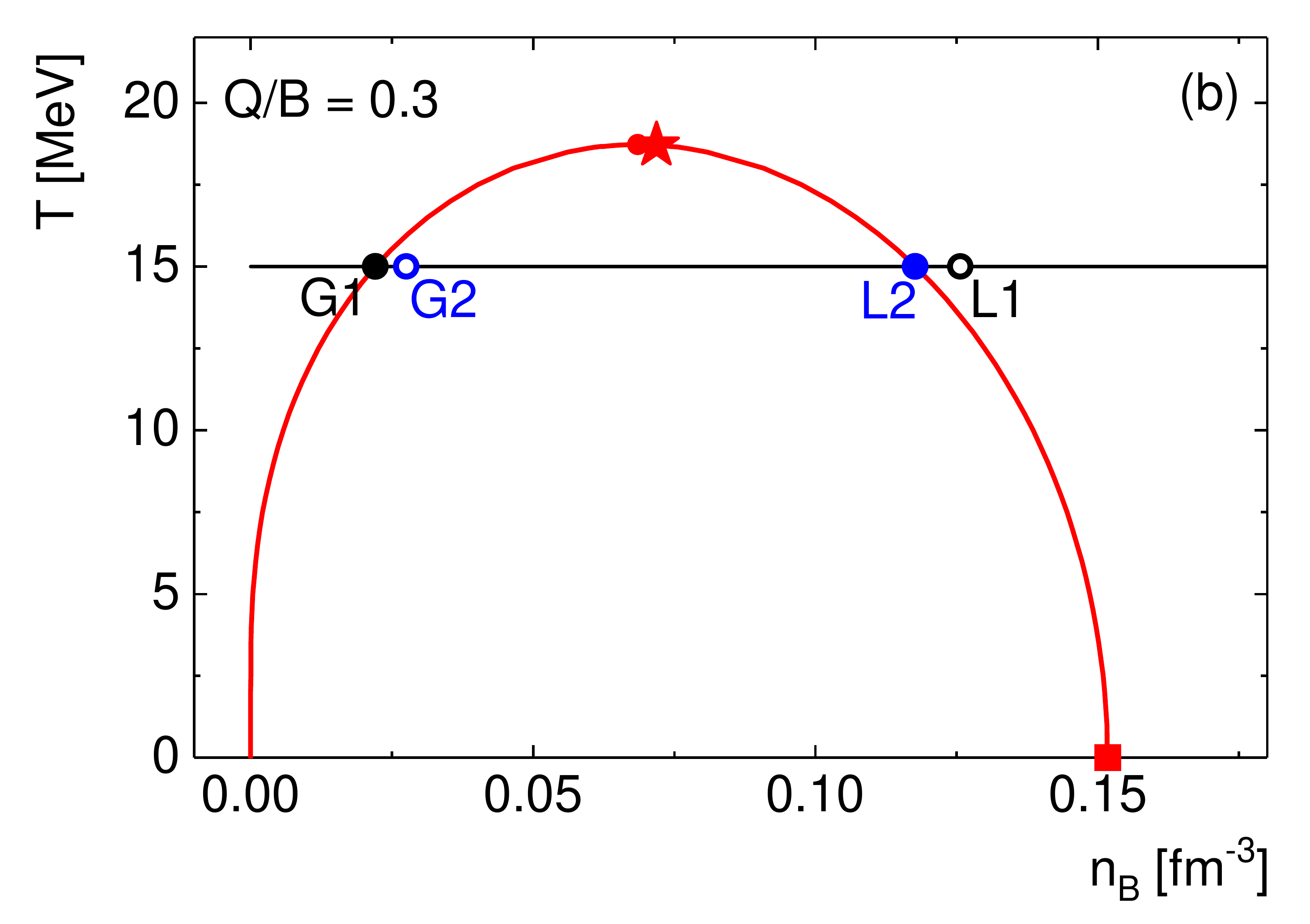}
\caption{\label{muT-2}  
Non-congruent LGPT regions ($a$) in the $(\mu_B, T)$ coordinates and ($b$) in the $(n_B, T)$ coordinates 
for the constant value $Q/B = 0.3$.
Critical point and temperature endpoint are represented by the stars and full circles, respectively.
Ground state is represented by the squares.
 Horizontal lines depict the isotherm at $Q/B=0.3$ and $T=15~ {\rm MeV}<T_c$.
}
\end{figure}
\begin{figure}[h!]
\includegraphics[width=0.49\textwidth]{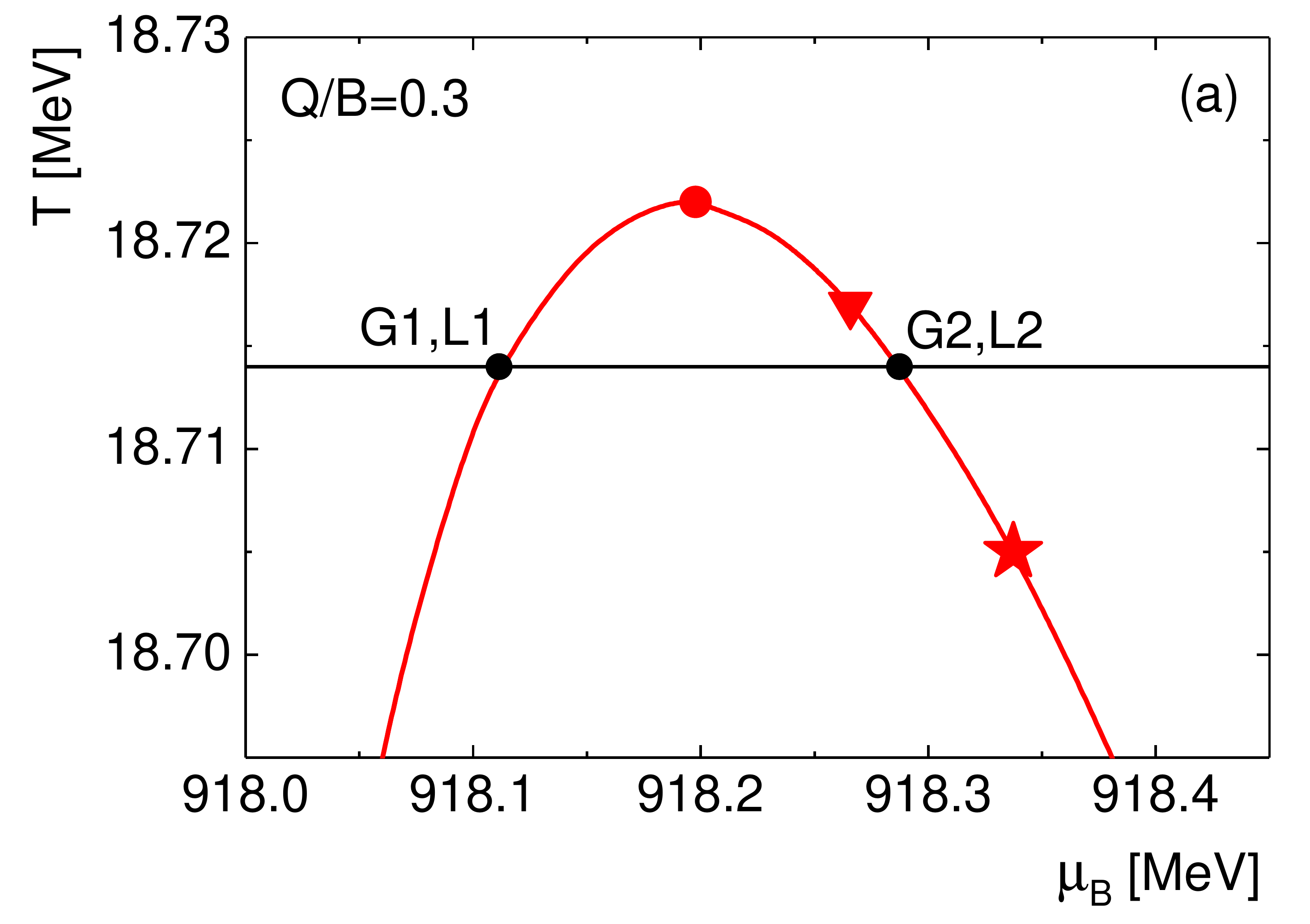}
\includegraphics[width=0.49\textwidth]{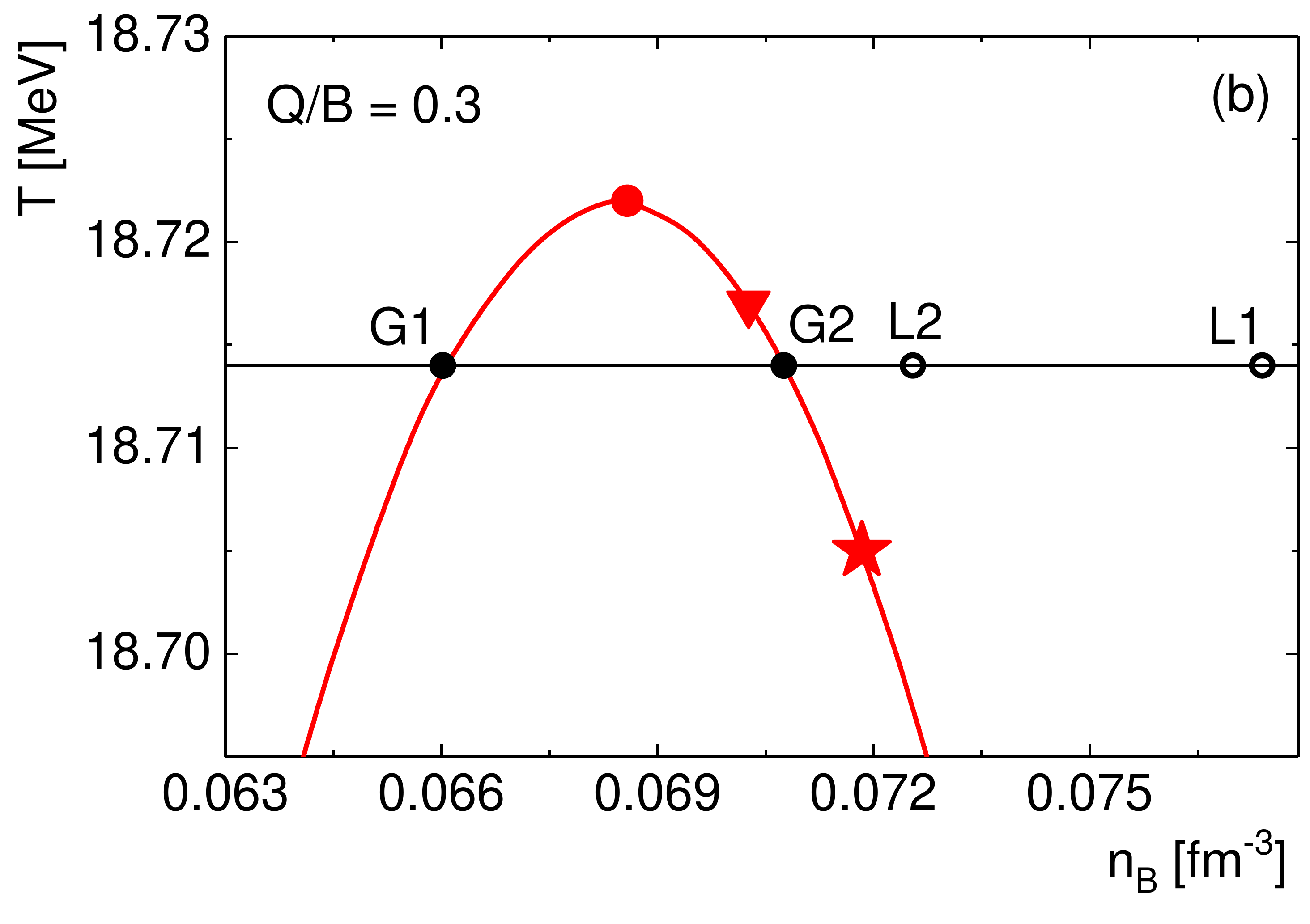}
\caption{\label{muT-zoom}  
The zoomed in picture of the CP region for the phase diagrams shown in Fig. \ref{muT-2}.
Horizontal lines depict the isotherm at $Q/B=0.3$ and $T_c<T=18.714~ {\rm MeV}<T_{TEP}$.
Pressure endpoint is represented by triangles.
}
\end{figure}

Figure~\ref{muT-2} presents an example of the phase transformation at a constant temperature, $T=15~ {\rm MeV}<T_c$, for the value of asymmetry parameter $Q/B=0.3$ by horizontal lines. 
At the constant temperature, $T<T_c$, the mixed phase of the non-congruent PT starts from a point
$G1=(T,\mu^{G1}_{B},\mu^{G1}_{Q})$, on the boiling curve of the phase diagram,
and finishes at a point $L1=(T,\mu^{L1}_{B},\mu^{L1}_{Q})$ on the saturation curve. The terms ``start to finish of the PT" denote a pictorial  progression -- from the small density (gas) to the large density (liquid).
The pure gaseous phase in the point $G1$ is in thermodynamical equilibrium with the infinitesimal liquid fraction in the point $L1$ while the pure liquid phase in the point $L2$ is in thermodynamical equilibrium with the infinitesimal gaseous fraction in the point $G2$. 
The locations of these four points $G1,~L1,~G2,~L2$ on the phase diagram at a given $T$ can be calculated from the Gibbs equilibrium conditions:
\eq{\label{gibbs-p}
p_{G1}(T,\mu^{G1}_{B},\mu^{G1}_{Q})~&=~p_{L1}(T,\mu^{G1}_{B},\mu^{G1}_{Q})~,\\
p_{L2}(T,\mu^{L2}_{B},\mu^{L2}_{Q})~&=~p_{G2}(T,\mu^{L2}_{B},\mu^{L2}_{Q})~.
}

While points ($G1$,$L1$) or ($L2$,$G2$) both have the same temperature, pressure, and chemical potentials, they differ by the values of the shifted chemical potential, $\mu^*_{B}$, and densities of charges, $n_{B}$, $n_Q$. The requirement (\ref{conserv-general}) of the constant charge ratio 
is imposed on the pure phases ($G1,L2$) only:
\eq{\label{conserv}
\frac{n_Q^{G1}(T,\mu^{G1}_{B},\mu^{G1}_{Q})}{n_B^{G1}(T,\mu^{G1}_{B},\mu^{G1}_{Q})}~=~\frac{n_Q^{L2}(T,\mu^{L2}_{B},\mu^{L2}_{Q})}{n_B^{L2}(T,\mu^{L2}_{B},\mu^{L2}_{Q})}~=~\frac{Q}{B}~=~{\rm const}~.
}

The infinitesimal fractions ($G2,L1$) are not restricted by this requirement, as their contributions
are infinitesimally small.
Here, we have four equations (\ref{gibbs-p})-(\ref{conserv}) for four unknown chemical potentials, $\mu^{G1}_{B}$, $\mu^{G1}_{Q}$, $\mu^{L2}_{B}$, and $\mu^{L2}_{Q}$. 
In contrast to the case of a congruent PT, both pressures and chemical potentials at the start and finish of liquidification are generally not identical, $p_{G1}\neq p_{L2}$, $\mu_B^{G1}\neq\mu_B^{L2}$. Thus, the dimensionality of the ($\mu_B$,$T$) phase diagram is increased.

The phase transformation starts from a gas in point G1 in equilibrium with an infinitesimal fraction of liquid in point L1. 
In the course of the phase transformation, the location of the liquid fraction moves from the point L1 towards the point L2, while the location the gaseous fraction moves from the point G1 towards the point G2. Phase transformation finishes with pure liquid in the point L2 in equilibrium with an infinitesimal fraction of gas in the point G2. Points G1, L1 and G2, L2 in the $(\mu_B, T)$-plane coincide, in accordance with the Gibbs condition. 

The mixed phase boundaries are found as the sets of point $G1$ (saturation curve) and $L2$ (boiling curve). Figure~\ref{muT-2} also presents the mixed phase boundaries for $Q/B=0.3$. 
Points L1, G2 which correspond to infinitesimal fractions of the mixed phase are not located at the correspondent mixed phase boundaries in the $(n_B, T)$-plane. While the point L1 is located in the pure liquid phase, the location of the point G2 is inside the mixed phase. Nevertheless, the point G2 corresponds to the pure infinitesimal gaseous fraction.

The standard equations (\ref{cp}) used for the determination of the CP in the congruent case are not valid in the case of a non-congruent PT. 
Therefore, we find the CP as the point on the mixed phase boundary where the two phases become identical, $\mu^{*G1}_{B}=\mu^{*L2}_{B}$, $n^{G1}_{B}=n^{L2}_{B}$. One sees that $T_c$ decreases when  $Q/B$ is decreased. For instance, at $Q/B=0.3$ the position of the CP is found to be:
\eq{
T_c~=~18.72~{\rm MeV}~,~~~~~n_c~=~0.068~{\rm fm^{-3}}~,~~~~~p_c~=~0.49~{\rm MeV~fm^{-3}}~.
}

\begin{figure}[h!]
\includegraphics[width=0.49\textwidth]{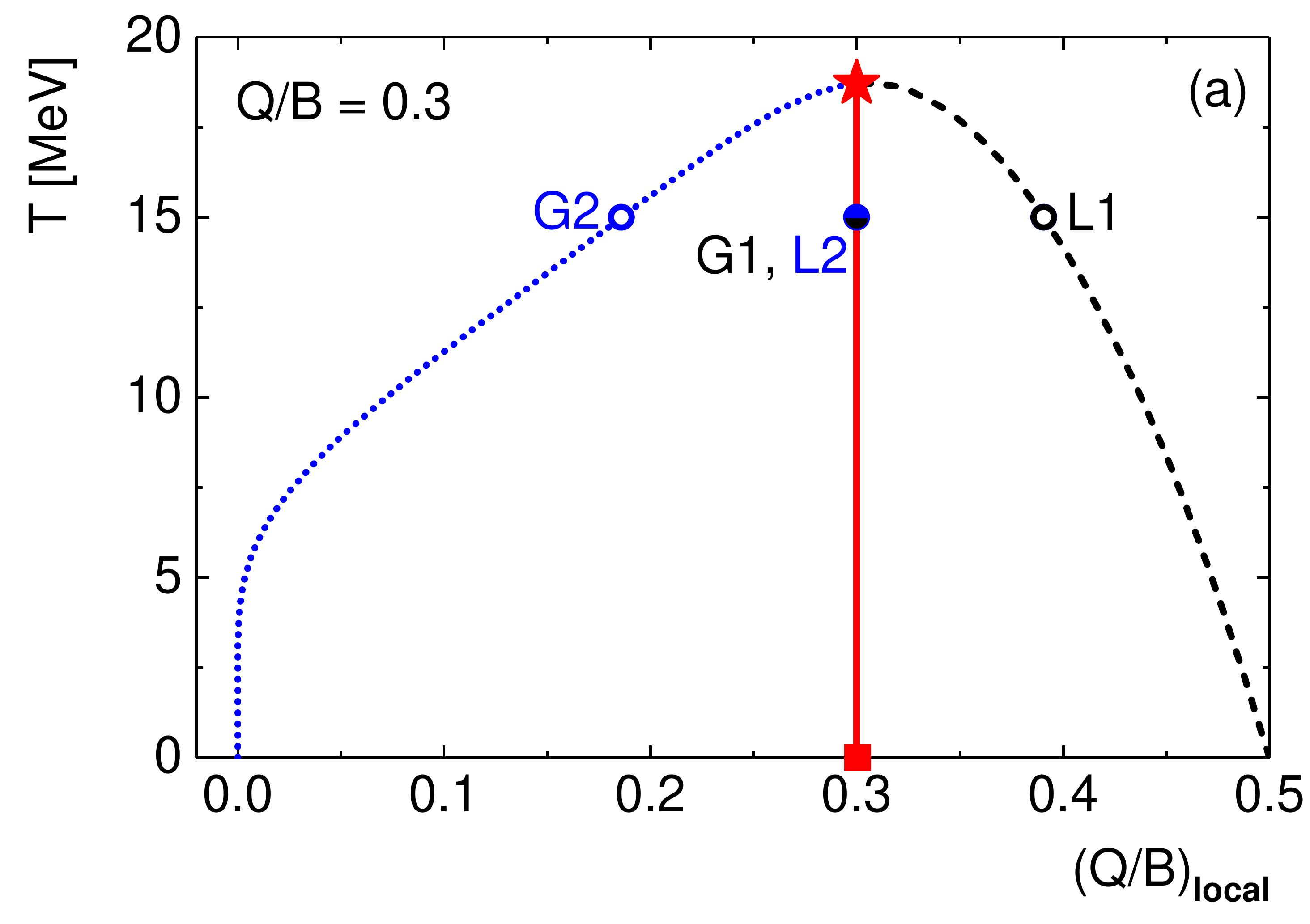}
\includegraphics[width=0.49\textwidth]{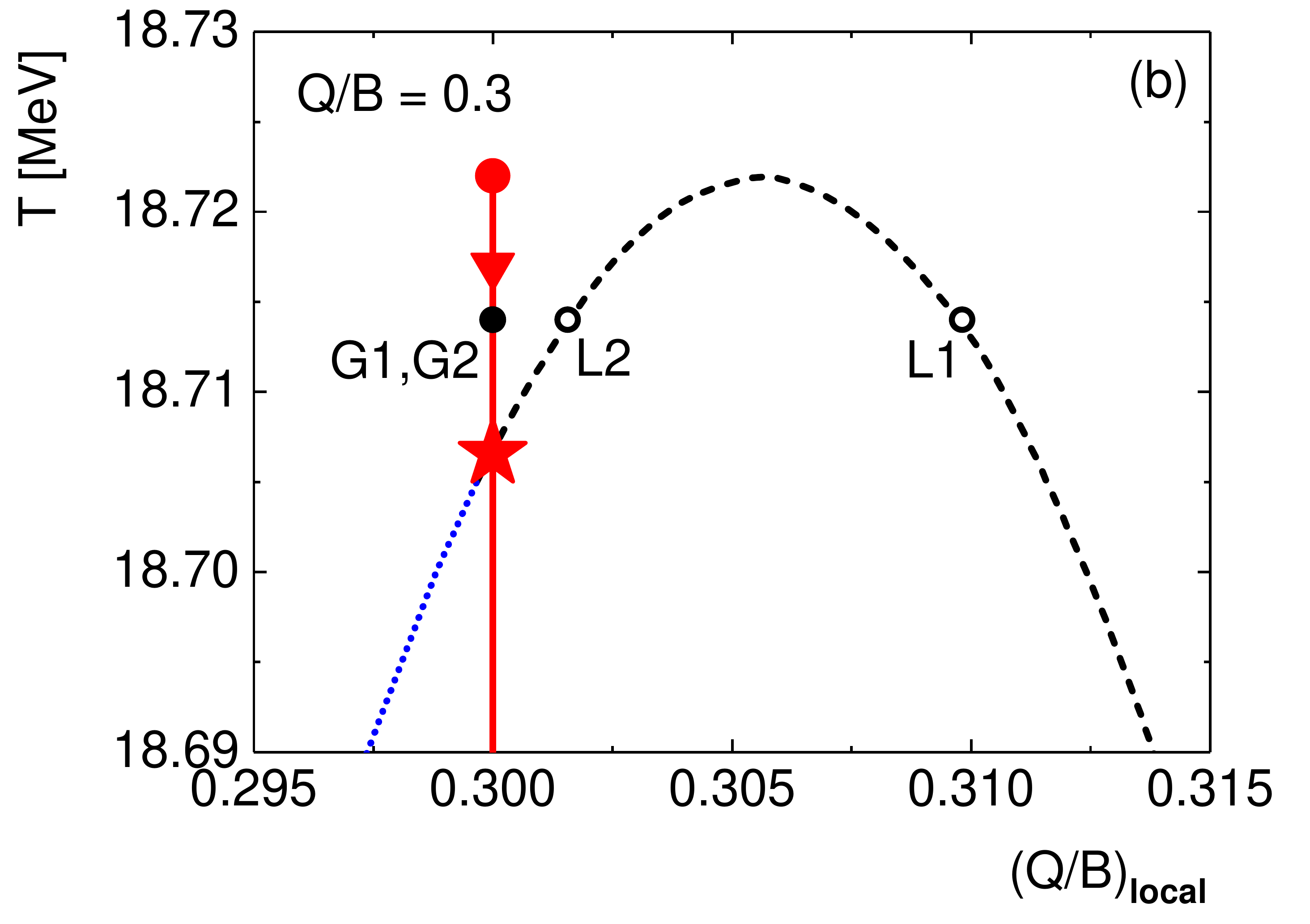}
\caption{\label{asymmetry}  $(a)$: Temperature of coexistence as a function of local charge fractions in coexistent phases, $(Q/B)_{local}$, for a fixed global charge fraction, $Q/B$=0.3. $(Q/B)_{local}=Q/B$ in pure phases on the borders of mixed phase are presented by thick red vertical line while $(Q/B)_{local}$ in correspondent liquid and gaseous infinitesimally small fractions are represented by dashed and dotted lines, respectively.  $(b)$: The zoomed in picture of the CP region. CP, TEP, pressure endpoint, and GS are shown by a star, a full circle, a triangle, and a square, respectively.
}
\end{figure}

\begin{figure}[h!]
\includegraphics[width=0.49\textwidth]{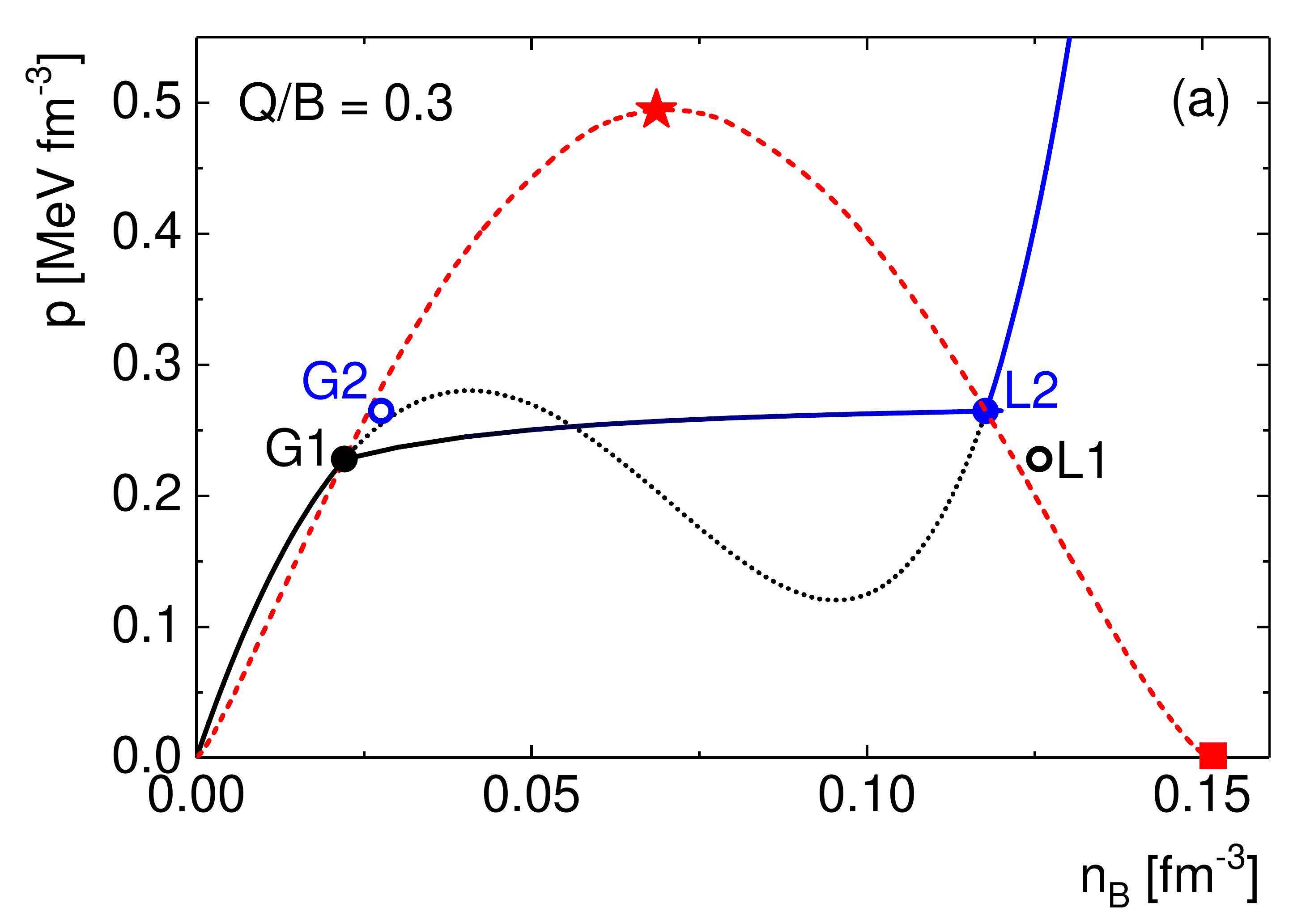}
\includegraphics[width=0.49\textwidth]{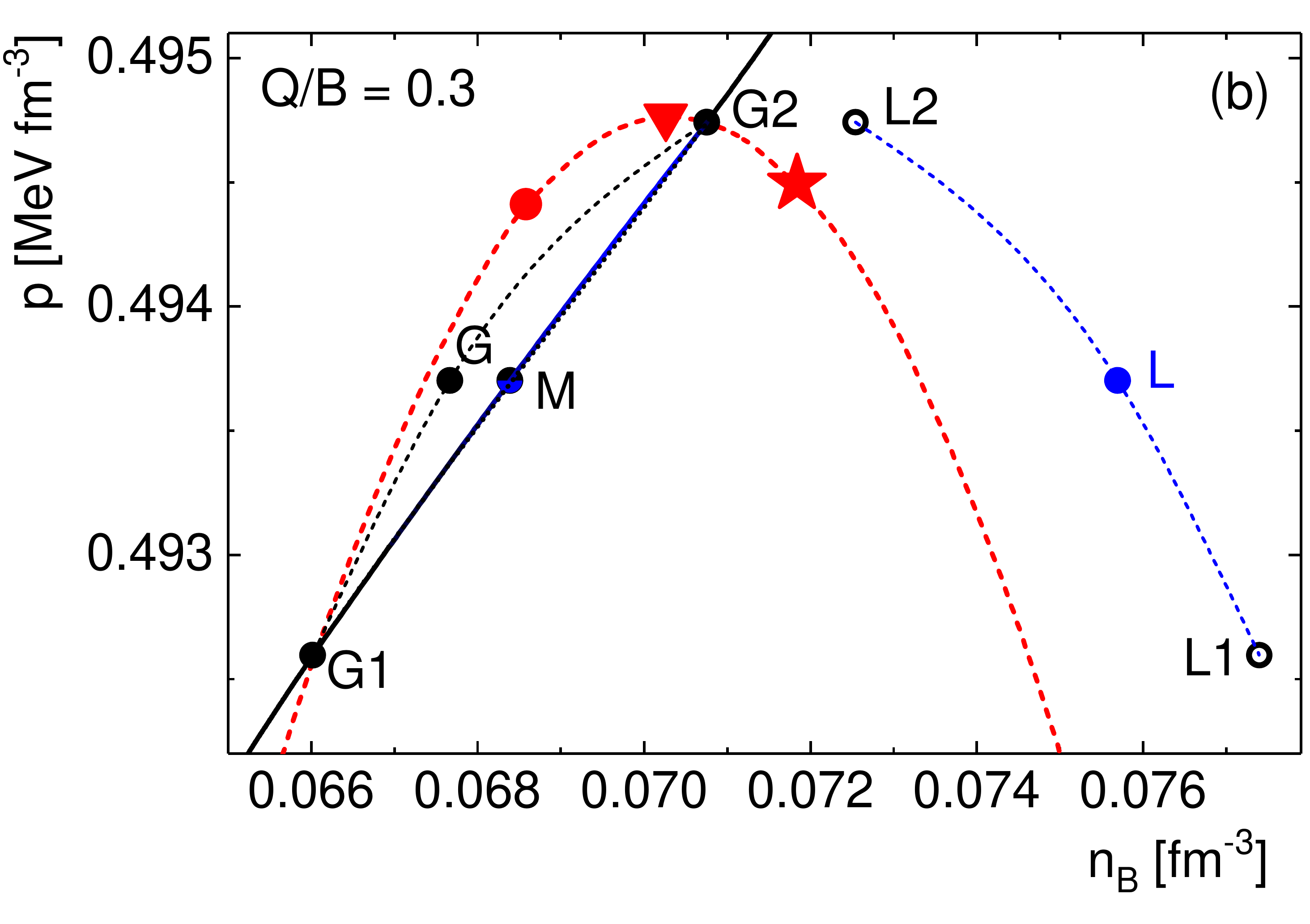}
\caption{\label{mixed_0_3} ($a$): $T=15~{\rm MeV}<T_c$ isotherm in the $(n_B, p)$ coordinates for a constant asymmetry parameter $Q/B=0.3$, solid curve. 
Meta-stable and unstable areas of isotherm are represented by dotted curve. Mixed phase boundaries are represented by red dashed curves.
($b$): The same as ($a$) but for the $T_c<T=18.714~{\rm MeV}<T_{TEP}$ isotherm.
The CP, the pressure endpoint, and the TEP, are shown by the star, the triangle, and the full circle, respectively.
The points G1,~L1,~G2,~L2 characterize the phase transformation along the isotherm, see text.
}
\end{figure}

\begin{figure}[h!]
\includegraphics[width=0.49\textwidth]{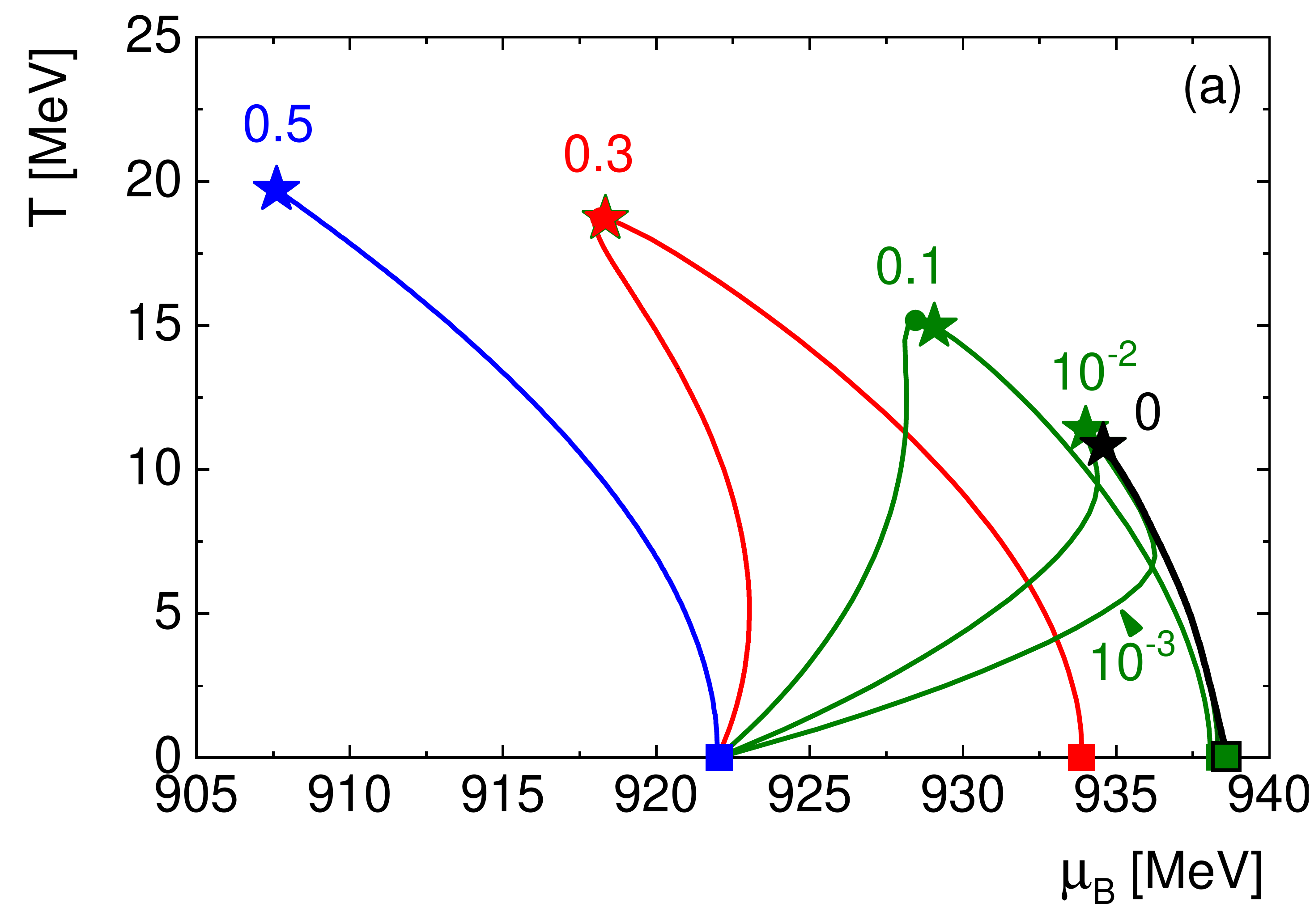}
\includegraphics[width=0.49\textwidth]{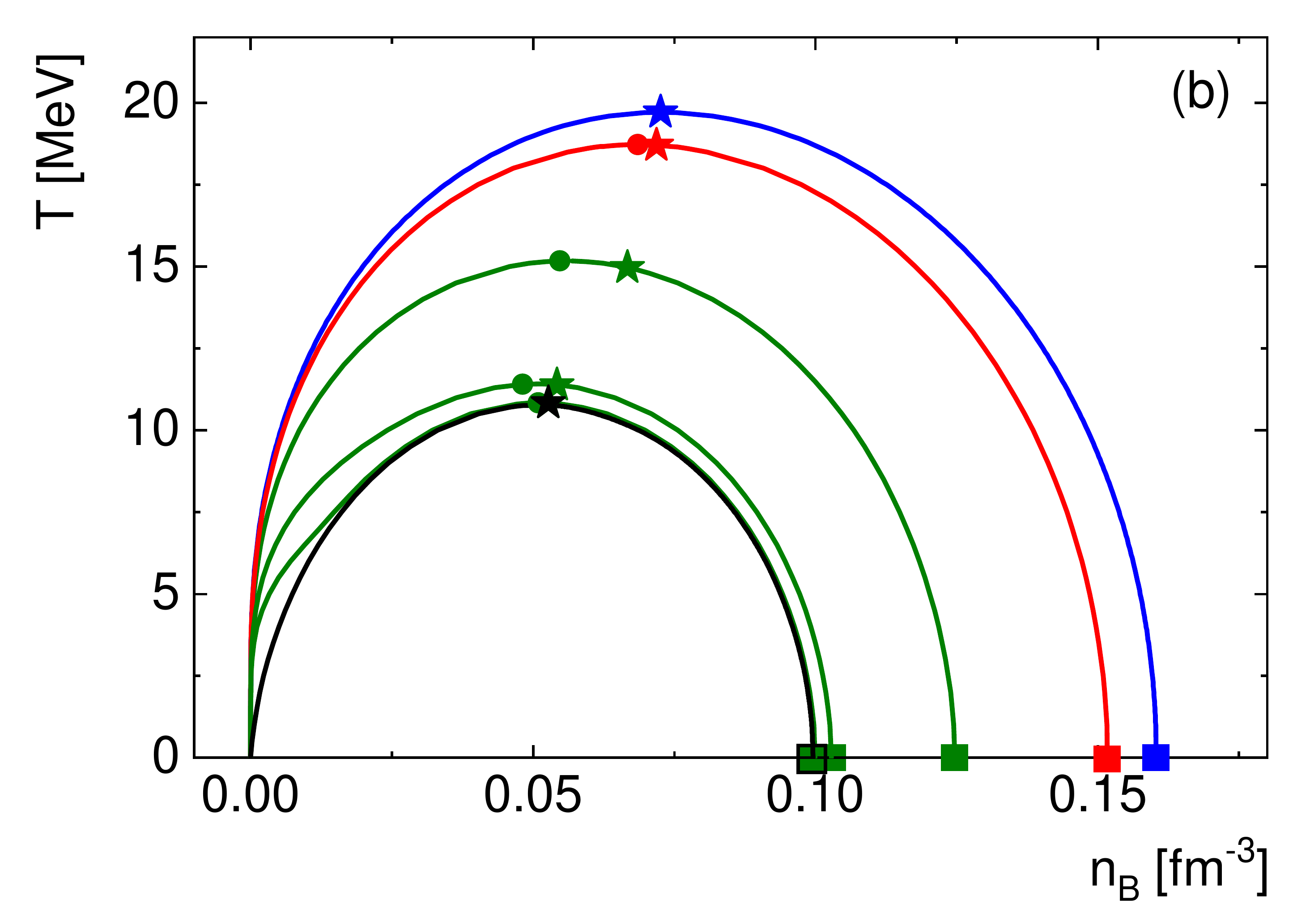}
\caption{\label{muT} Congruent and non-congruent LGPT regions in the $(\mu_B, T)$ coordinates ($a$) and in the $(n_B, T)$ coordinates ($b$)
for the constant $Q/B$ values.
Solid blue, red, and black lines correspond to $Q/B=0.5$, $0.3$, and $0$ respectively. Green lines correspond to $Q/B=0.1,10^{-2},10^{-3}$.
Critical points and temperature endpoints are represented by stars and full circles, respectively.
Ground states for different $Q/B$ are represented by squares.
}
\end{figure}

Another interesting feature of non-congruent PTs is that the locations of temperature and pressure endpoints (the point with, respectively, maximum $T$ and $p$ at which the phase coexistence is possible) differ from each other and from the location of the CP.
Temperature endpoints (TEPs) are shown by the full circles in Figs.~\ref{muT-2}-\ref{mixed_0_3},\ref{muT}. 
For all $Q/B$ values one finds $T_{\rm TEP}\geq T_c$.
The same is true for the temperature in the pressure endpoints.
Pressure endpoint for one value of $Q/B=0.3$ is shown by the triangles in Figs.~\ref{muT-zoom}-\ref{mixed_0_3}.

Figure \ref{muT-zoom} presents an example of the phase transformation at a constant temperature, 
$T_c<T=18.714~{\rm MeV}<T_{TEP}$, for the value of asymmetry parameter $Q/B=0.3$ by horizontal lines. 
At the constant temperature, $T_c<T<T_{TEP}$, the mixed phase of the non-congruent PT starts from a point
$G1=(T,\mu^{G1}_{B},\mu^{G1}_{Q})$, on the boiling curve of the phase diagram,
and finishes at a point $G2=(T,\mu^{G2}_{B},\mu^{G2}_{Q})$ on the boiling curve. Gaseous fractions in points $G1,G2$ are in Gibbs equilibrium with infinitesimal liquid fractions in points $L1,L2$, respectively.

Figure~\ref{asymmetry} shows the local nuclear matter charge fractions, $(Q/B)_{local}$, in the coexistent phases on the borders of mixed phase region for a fixed global charge fraction, $Q/B=0.3$. The black dashed curve represents $(Q/B)_{local}$ of the infinitesimal liquid fraction at the start of the mixed phase. The blue dotted curve represents $(Q/B)_{local}$ of the infinitesimal gaseous fraction at the finish of the mixed phase. The solid red line represents $(Q/B)_{local}=Q/B=0.3$ of both, the pure gaseous and the pure liquid phases, respectively, at the start and at the finish of the phase transformation.

The asymmetry in the gaseous fraction of the mixture is always larger then the asymmetry in the liquid fraction: this is the isospin distillation phenomenon, which is just the equivalent of the strangeness distillary considered in Refs. \cite{Greiner:1987tg,Greiner:1988pc,Greiner:1991us,Glendenning:1992vb}.
At $T\rightarrow 0$, the infinitesimal gaseous fraction approaches the composition of the pure neutron matter, $(Q/B)_{local}\rightarrow 0$, while the infinitesimal liquid fraction approaches the composition of the symmetric nuclear matter, $(Q/B)_{local}\rightarrow 0.5$.
Figure~\ref{asymmetry} $(b)$ shows a zoomed picture of the CP region.
 The asymmetry of the infinitesimal gaseous fraction becomes equal to the global asymmetry, $Q/B=0.3$, at the critical temperature, $T=T_c$.

The system can be described in the mixed phase for an arbitrary proportion of fractions by introducing an additional parameter $\chi$ -- the share of the total volume which is occupied by the liquid fraction. $\chi=0$ and $\chi=1$ correspond to the purely gaseous and to the purely liquid phase, respectively, while $0<\chi<1$ is realized for the mixed phase. The densities of the baryonic and of the electric charge for both, pure phases and the mixed phase, are given by
\eq{\label{nB-mixed}
n_B(T,\mu_{B},\mu_{Q})~=~(1~-~\chi)~n_B^G(T,\mu_{B},\mu_{Q})~+~\chi~n_B^L(T,\mu_{B},\mu_{Q})~,\\
n_Q(T,\mu_{B},\mu_{Q})~=~(1~-~\chi)~n_Q^G(T,\mu_{B},\mu_{Q})~+~\chi~n_Q^L(T,\mu_{B},\mu_{Q})~.
}
Here $n_B^G$, $n_Q^G$ and $n_B^L$, $n_Q^L$ are the charge densities of the gaseous and the liquid fractions, respectively.
In general, the volumes occupied by both fractions are finite. Hence, conservation laws shall be applied to the total mixture only, but not to the different components separately \cite{Greiner:1987tg,Greiner:1988pc,Greiner:1991us,Glendenning:1992vb}. The requirement of a constant charge ratio (\ref{conserv-general}) for both, pure phases and the mixed phase, reads
\eq{\label{conserv-mixed}
\frac{(1~-~\chi)~n_{Q}^G(T,\mu_{B},\mu_{Q})~+~\chi~n_{Q}^L(T,\mu_{B},\mu_{Q})}{(1~-~\chi)~n_B^G(T,\mu_{B},\mu_{Q})~+~\chi~n_B^L(T,\mu_{B},\mu_{Q})}~=~\frac{Q}{B}~=~{\rm const}~.
}
 The chemical potentials $\mu_{B}$ and $\mu_{Q}$ can be found at constant $T$ and $\chi$ from equation (\ref{conserv-mixed}) evaluated simultaneously with the Gibbs equilibrium condition,
\eq{
p_G(T,\mu_{B},\mu_{Q})~=~p_L(T,\mu_{B},\mu_{Q})
}

Figure~\ref{mixed_0_3} ($a$) shows the $T=15~{\rm MeV}<T_c$ isotherm in $(n_B, p)$ coordinates for $Q/B=0.3$, including the mixed phase region. Points G1, L1, G2, L2 are also presented. The solid curve represents the stable solution, while dotted curve represents the meta-stable and unstable areas.
Note that the pressure is not constant in the mixed phase -- a generic feature of all non-congruent PTs. This is due to the presence of  additional degrees of freedom in the mixed phase -- here the local asymmetry parameter $(Q/B)_{local}$.
Figure~\ref{mixed_0_3} ($b$) is the same as Fig.~\ref{mixed_0_3}($a$) but for the supercritical isotherm $T_c<T=18.714~{\rm MeV}<T_{TEP}$. Point M shows the example of the mixed phase which includes gaseous, point G, and liquid, point L, fractions in Gibbs equilibrium. Black and blue dashed lines represent trajectories of, respectively, gaseous and liquid fractions in the mixture from start to finish of the phase transition.
\begin{figure}[h!]
\includegraphics[width=0.70\textwidth]{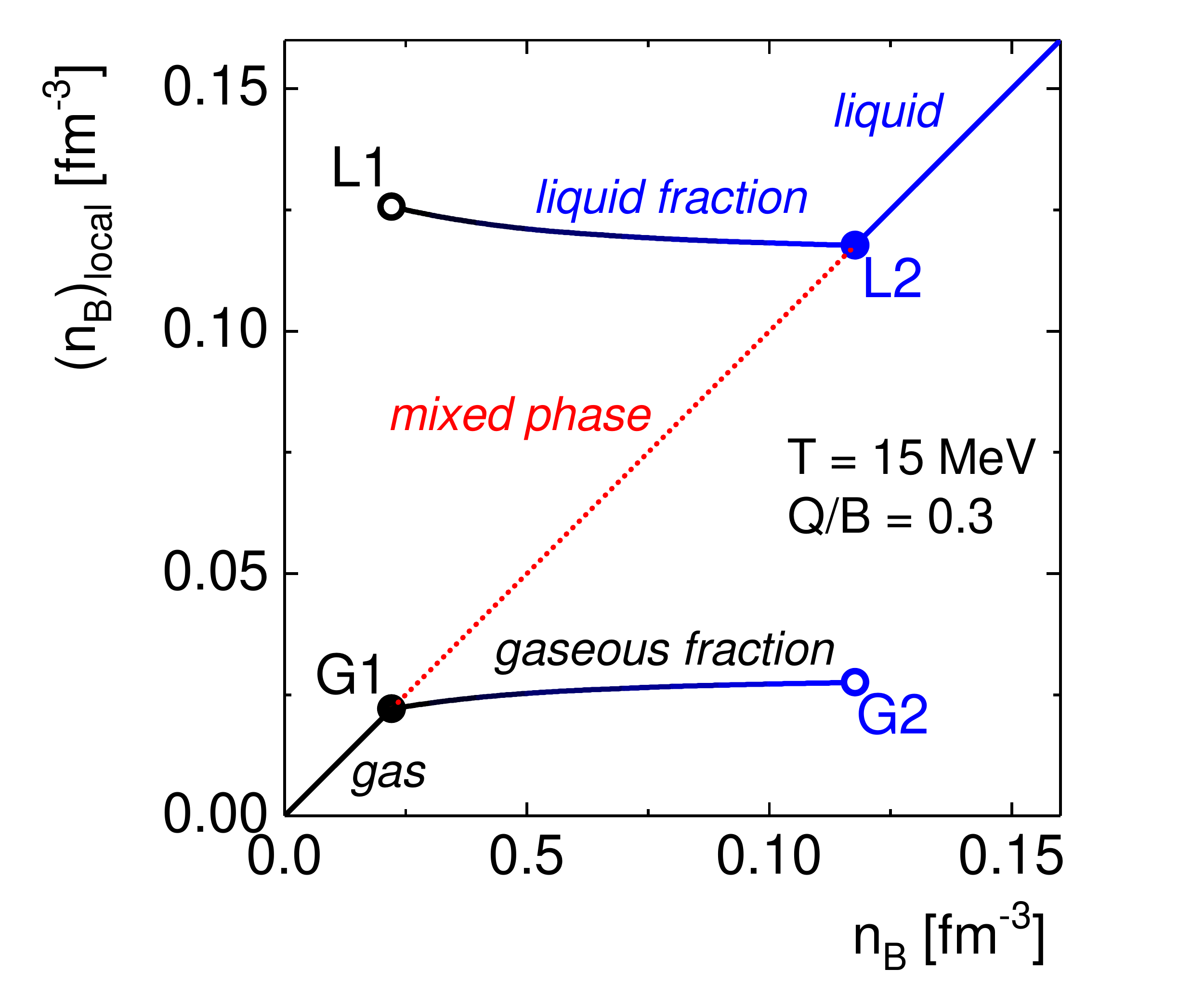}
\caption{\label{nB_local} Local baryon densities in the gaseous and in the liquid fractions of the mixed phase as functions of the total baryon density for $Q/B=0.3$ and $T=15~{\rm MeV}$.
}
\end{figure}

Figure~\ref{nB_local} shows the local baryon densities, $(n_B)_{local}$, in the gaseous and in the liquid fractions of the mixed phase as functions of the total baryon density, $n_B$, for $Q/B=0.3$ and $T=15~{\rm MeV}<T_c$. 
At $n_B<n^{G1}_B$ one has the pure gaseous phase only, whereas at $n_B>n^{L2}_B$ the pure liquid phase is realized.

\begin{figure}[h!]
\includegraphics[width=0.55\textwidth]{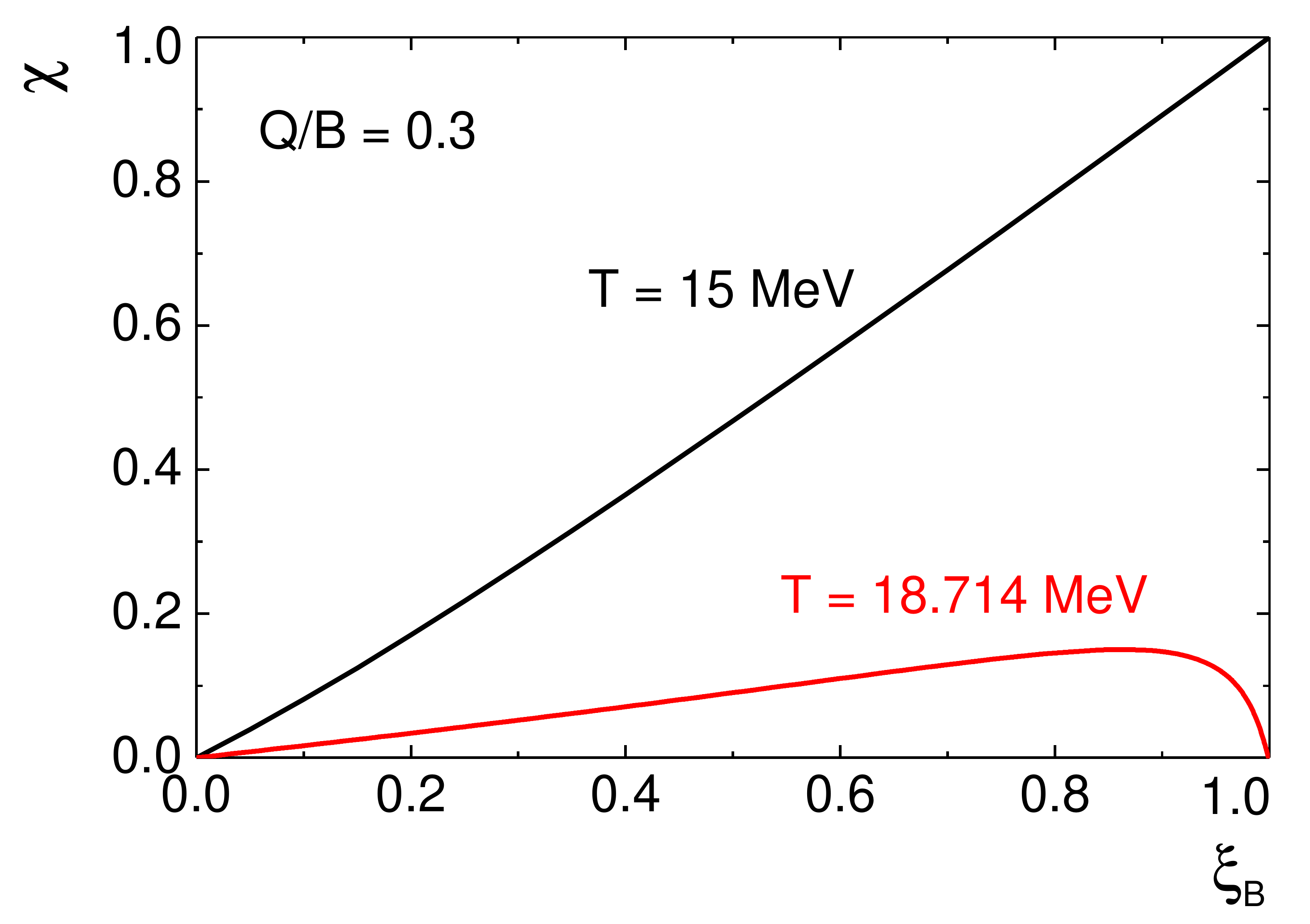}
\caption{\label{fig-chi}  Fraction of volume occupied in the mixed phase by liquid, $\chi$, as a function of baryon density for global asymmetry $Q/B=0.3$. As a measure of density $\xi_B$ is taken. For all $T$ the start and the finish of the PT correspond to $\xi_B=0$ and $\xi_B=1$, respectively. The black curve corresponds to $T=15~{\rm MeV}<T_c$ while the red curve corresponds to $T_c<T=18.714~{\rm MeV}<T_{\rm TEP}$.
}
\end{figure}

Figure~\ref{fig-chi} shows $\chi$ as a function of the baryon density in the mixed phase. The measure of the density for a fixed $T$ is taken as $\xi_B=(n_B-n^{G1}_B)/(n^{L2}_B-n^{G1}_B)$. 
The quantity $\xi_B$ is constructed in such a way that, for all $T$, $\xi_B=0$ at the start of the PT and $\xi_B=1$ at the finish of the PT. $Q/B=0.3$ is used in Fig.~\ref{fig-chi}. The two lines correspond to temperatures $T=15~{\rm MeV}<T_c$ and $T_c<T=18.714~{\rm MeV}<T_{\rm TEP}$. The fraction of the liquid is zero in both cases at the start of the PT. The liquid fraction starts to increase monotonically when the density increases. 
In the case of subcritical temperatures, $T<T_c$, the liquid fraction reaches $\chi=1$ at the finish of the PT. Correspondingly, the  gaseous fraction reaches $1-\chi=0$, i.e., only the pure liquid phase is left. In contrast, the so-called \emph{retrograde~condensation}~\cite{kuenen1892retrograde,hicks1975gas} occurs for supercritical temperatures in the narrow temperature range, $T_c<T<T_{\rm TEP}$:  at some value of the density, $n_B^{\rm max}$, the fraction of the liquid phase reaches its maximum value, $\chi_{\rm max}$, and with further increase of density, decreases rapidly. No liquid remains at the finish of the PT, and the system is in the purely gaseous phase again. Retrograde condensation is a unique feature of non-congruent PTs \cite{Huang_1963}.

Figure~\ref{muT} ($b$) presents the LGPT regions obtained for constant $0 \leq Q/B \leq 0.5$ values. Critical points are shown by the stars. The picture for $0.5 \leq Q/B \leq 1$ mirrors the $0 \leq Q/B \leq 0.5$ picture, as follows from the  isospin symmetry in nuclear matter.
The $Q/B=10^{-3}$ and $Q/B=0$ mixed phase boundaries in the ($n_{B}$, $T$) coordinates virtually coincide except for low densities.  
The location of the start of the mixed phase, at $T=0$, is independent of $Q/B$, for all non-zero $Q/B$ values. 
This point corresponds to the zero nucleon density. 
The non-congruence of PT disappears for all non-zero finite values of $T$ as $Q/B\rightarrow 0$. The LGPT region in ($\mu_B$,~$T$) plane shrinks and finally degenerates to the LGPT line  and $\{T_{\rm TEP},\mu_{\rm TEP}\}\rightarrow\{T_c,\mu_c\}\rightarrow\{T_c,\mu_c\}_{Q/B=0}$. However, the non-congruence of the PT remains at $T=0$.

\section{Fluctuations of baryonic and electric charges}
\label{sec:fluct}

The scaled variances of baryonic and electric charges fluctuations can be calculated as
\eq{
\omega[B]&=\frac{\mean{B^2}-\mean{B}^2}{\mean{B}}=\frac{T}{n_B}\left[\frac{\partial^2 p}{(\partial \mu_B)^2}\right]_{T,\mu_Q}{\rm and}~~\omega[Q]=\frac{\mean{Q^2}-\mean{Q}^2}{\mean{Q}}=\frac{T}{n_Q}\left[\frac{\partial^2 p}{(\partial \mu_Q)^2}\right]_{T,\mu_B},
}
respectively. This yields the following expressions:
\eq{\label{wb}
\omega[B]&=\omega^{id}[B]\left[\frac{1}{(1-b n_B)^{2}}-\frac{2 a n_B}{T} \omega^{id}[B]\right]^{-1}~,\\\label{wq}
\omega[Q]&=\ddfrac{b^2 n_N n_Q \omega^{id}[N]+\left[(1-bn_Q)^2-\frac{2 a n_N}{T} (1-b n_B)^2 \omega^{id}[N]\right]\omega^{id}[Q]}{1-\frac{2 a n_B}{T} (1-b n_B)^2\omega^{id}[B]}~.
}
Here $n_N=n_B-n_Q$ is the neutrons number density. The ideal gas scaled variances of the baryon, neutron, and proton multiplicity fluctuations are given by
\eq{\nonumber
\omega^{id}[B]&=\frac{{\rm Var}^{id}_n(T,\mu^*_B)+{\rm Var}^{id}_p(T,\mu^*_B+\mu_Q)}{n^{id}_n(T,\mu^*_B)+n^{id}_p(T,\mu^*_B+\mu_Q)}~,\\\label{w-id}
\omega^{id}[N]&=\frac{{\rm Var}^{id}_n(T,\mu^*_B)}{n^{id}_n(T,\mu^*_B)}~,~~~~~~~{\rm and}~~~~~~~
\omega^{id}[Q]=\frac{{\rm Var}^{id}_p(T,\mu^*_B+\mu_Q)}{n^{id}_p(T,\mu^*_B+\mu_Q)}~,
}
respectively. The ideal gas variance of particles multiplicity fluctuations is calculated as
\eq{
{\rm Var}^{id}_j(T,\mu_j)=\left[\frac{\partial^2 p^{id}_j(T,\mu_j)}{(\partial \mu_j)^2}\right]_{T}=n^{id}_j(T,\mu_j)-\frac{ g}{2\pi^2} \int_0^{\infty} k^2\,dk
\, f^2_{\rm k}(T,\mu_j)~.
}
For symmetric nuclear matter, $\mu_Q=0$, Eq.~(\ref{wq}) reduces to the following form,
\eq{
\omega[Q]
&=\omega^{id}[Q] \ddfrac{1-b n_B+\frac{b^2 n_B^2}{2}-\frac{ a n_B}{T} (1-b n_B)^2\omega^{id}[Q]}{1-\frac{2 a n_B}{T} (1-b n_B)^2\omega^{id}[Q]}~,
}
where $\omega^{id}[Q]=\omega^{id}[N]=\omega^{id}[B]$. Thus, even in the symmetric case, $\omega[Q]$ is not a linear function of $\omega[B]$, due to the neutron-proton correlations.

From Eq.~(\ref{w-id}) it follows that, in the Boltzmann approximation, $\omega^{id}[B]=\omega^{id}[Q]=\omega^{id}[N]=1$ and Eqs.~(\ref{wb}),(\ref{wq}) reduce to
\eq{\label{wb-qvdw-boltz}
\omega[B]=\left[\frac{1}{(1-b n_B)^{2}}-\frac{2 a n_B}{T} \right]^{-1},
~~~~~
\omega[Q]=\ddfrac{b^2 n_N n_Q +(1-bn_Q)^2-\frac{2 a n_N}{T} (1-b n_B)^2 }{1-\frac{2 a n_B}{T} (1-b n_B)^2}~.
} 
\begin{figure}[h!]
\includegraphics[width=0.49\textwidth]{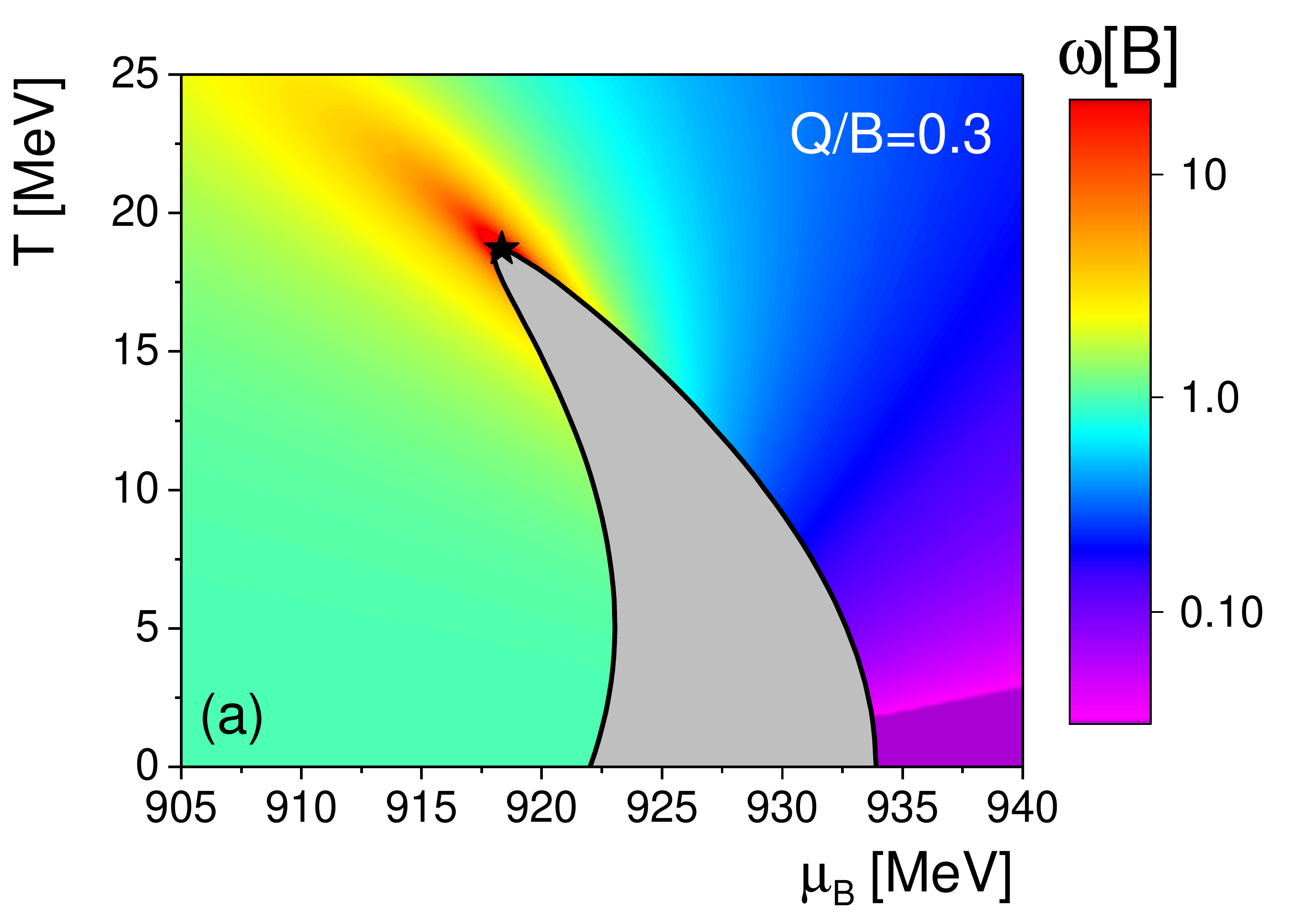}
\includegraphics[width=0.49\textwidth]{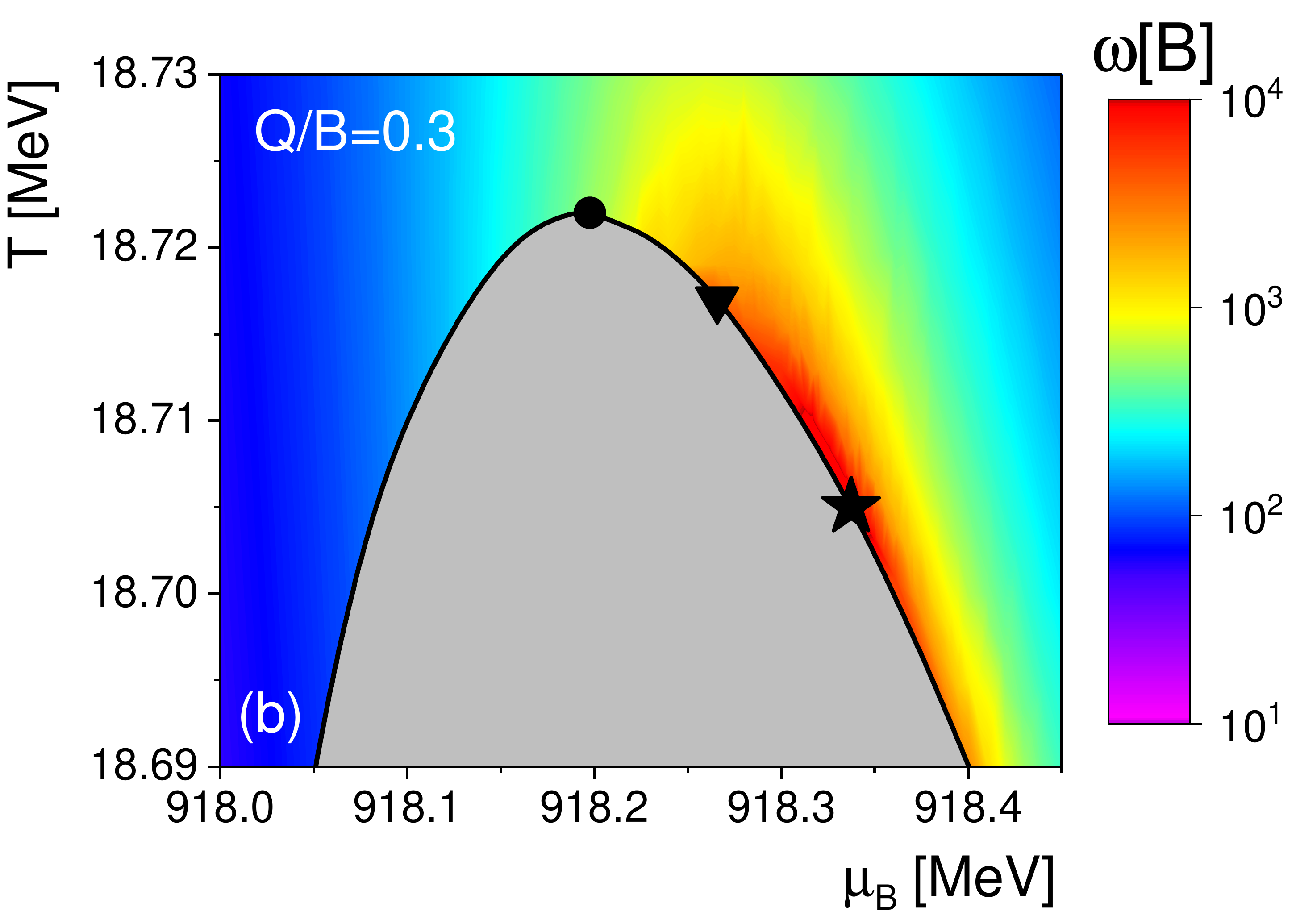}
\includegraphics[width=0.49\textwidth]{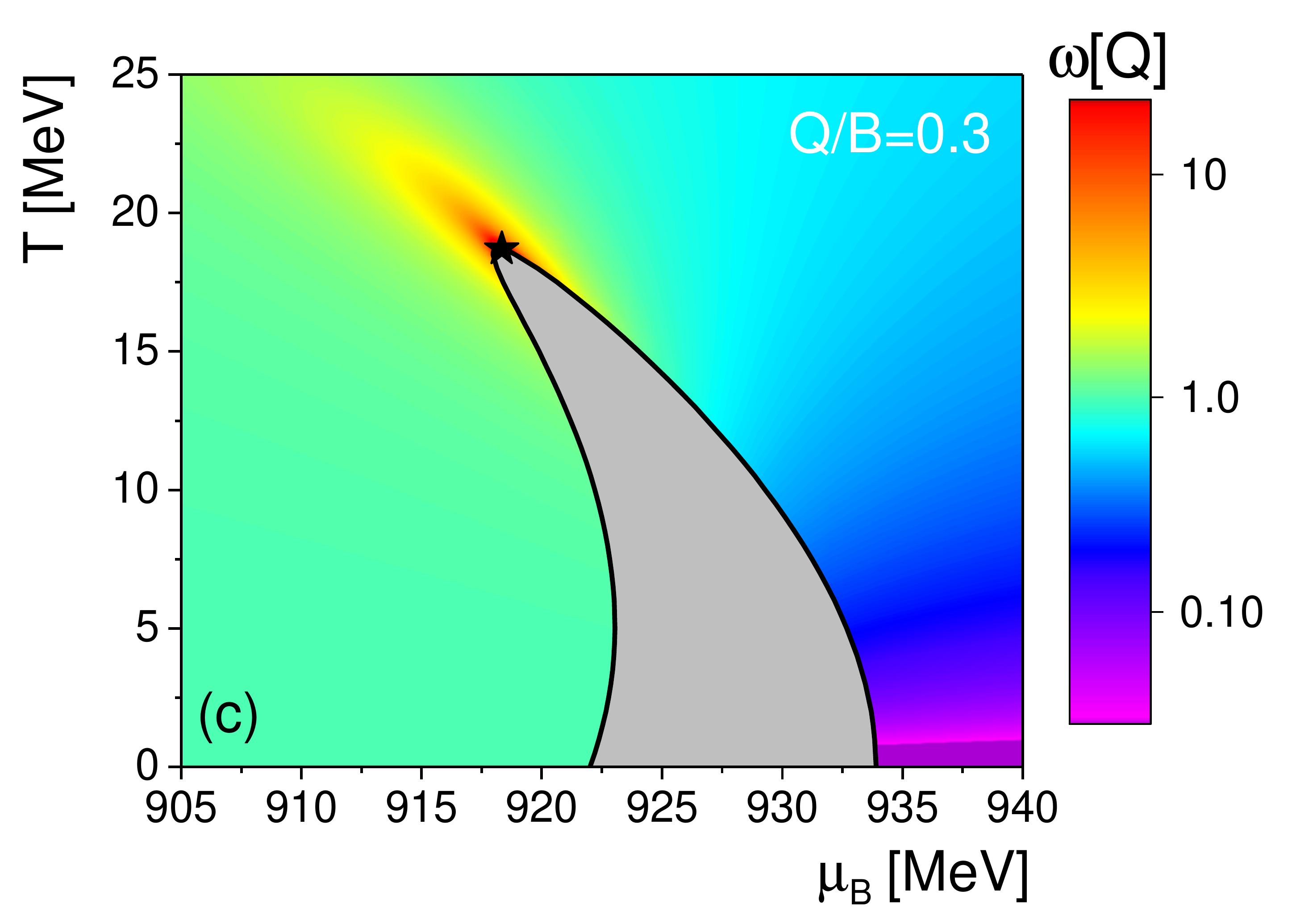}
\includegraphics[width=0.49\textwidth]{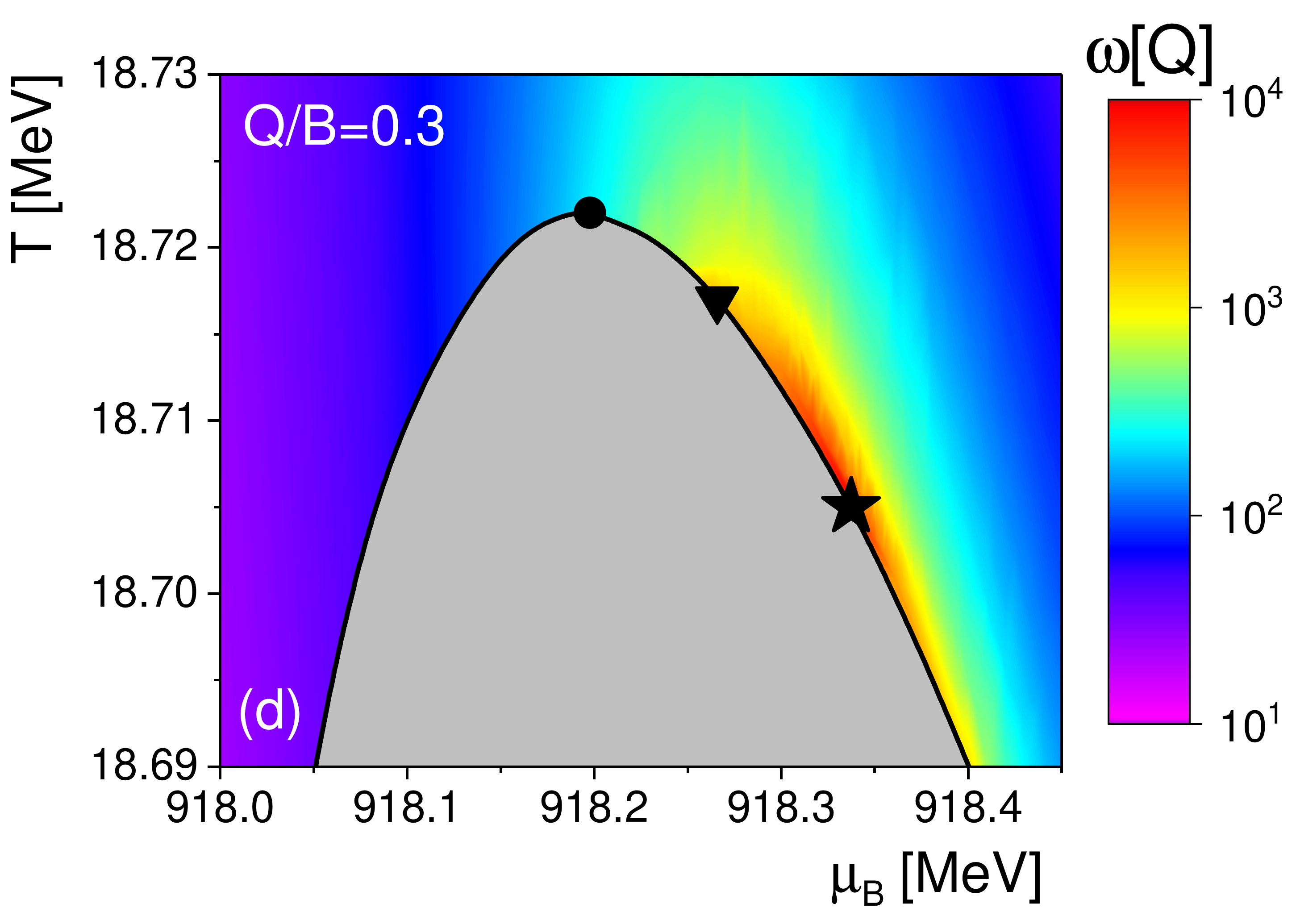}
\caption{\label{flukes}  The scaled variance of the baryonic ($a$,$b$) and of the electric ($c$,$d$) charge in the $(\mu_B, T)$ coordinates for asymmetric nuclear matter with the asymmetry parameter $Q/B=0.3$. 
The critical point, the temperature endpoint, and the pressure endpoint are represented by stars, full circles, and triangles respectively.  The zoomed in picture of the CP region is shown in ($b$) and ($d$).
}
\end{figure}

The scaled variances given by Eq.~(\ref{wb-qvdw-boltz}) satisfy the relation
\eq{\label{q}
\omega[Q]~=~q~\omega[B]~+~1~-~q~,
}
where $q=n_Q/n_B$ is the probability of detecting a proton, i.e., the charge fluctuations are obtained from binomial folding of baryon number fluctuations. This is a well-known generic feature of Boltzmann systems \cite{Begun:2004gs}. In contrast, the quantum statistics does violate the relation (\ref{q}).

The correlation between the baryonic and the electric charges is calculated as follows:
\eq{\nonumber
\frac{{\rm cor[B,Q]}}{V}&=\frac{\mean{B Q}-\mean{B}\mean{Q}}{V}=
T \left[\frac{\partial}{\partial \mu_Q}\right]_{T,\mu_B}\left[\frac{\partial p}{\partial \mu_B}\right]_{T,\mu_Q}\\
&=\frac{n_Q}{1-b n_B}\left[\frac{\omega^{id}[Q]}{\omega^{id}[B]}-b n_B\right]\omega[B]~.
}

Figure~\ref{flukes} presents the scaled variances of baryonic, $\omega[B]$, and electric, $\omega[Q]$,  charges fluctuations in the ($\mu_B$, $T$) coordinates for the region of the pure phases. Generally, $\omega[B]>\omega[Q]$ in every point of the ($\mu_B, T$)-plane. At $T\rightarrow 0$, both $\omega[B]$ and $\omega[Q]$ approach unity in the gaseous phase (Poisson distributions) and zero in the liquid phase (dense packing limit). Both $\omega[B]$ and $\omega[Q]$  are divergent at the CP. At the same time, both these quantities exhibit regular behavior at the TEP of the non-congruent LGPT.

\section{Isospin-dependent interaction parameters}
\label{sec:isospin-dep}

Throughout this work we have assumed that the QvdW interaction parameters are the same for proton-proton, proton-neutron, and neutron-neutron interactions, i.e. the isospin dependence of the $NN$ potential was not considered.
The model predicts the symmetry energy value of about $J \simeq 20$~MeV, which is 10-15~MeV lower than the empirical estimate~\cite{Stone:2006fn}.
In addition, the liquid-gas phase transition in pure neutron matter, predicted by the model~(see Fig.~\ref{congr}), appears to be ruled out by chiral effective field theory~\cite{Hebeler:2013nza}.
An improved description of asymmetric nuclear matter within the QvdW approach can be achieved by considering isospin dependent QvdW parameters.

In Ref.~\cite{Vovchenko:2017zpj} the multi-component QvdW model was formulated where one can specify the attractive and repulsive QvdW parameters for each pair of particle species.
This formalism is applied here for asymmetric nuclear matter. 
We define $a_{nn}$, $a_{pp}$, and $a_{pn}=a_{np}$ as attractive QvdW parameters for neutron-neutron, proton-proton, and neutron-proton interactions, respectively.
Similarly, $b_{nn}$, $b_{pp}$, and $b_{pn}=b_{np}$ are repulsive QvdW parameters.
The assumed isospin symmetry yields $b_{pp} = b_{nn}$ and
 $a_{pp} = a_{nn}$.
Therefore, the model has four QvdW interaction parameters: $a_{nn}$, $a_{np}$, $b_{nn}$,  and $b_{pn}$. 
The isospin-dependent multi-component QvdW equation for the pressure 
is 
\begin{eqnarray}
 p(T,\mu_B,\mu_Q) ~=~ p_{n}^{\rm id}( T,\mu_n^*)~+~ p_{p}^{\rm id}( T,\mu_p^* )~-~a_{nn}~(n^2_n~+~n^2_p)~-~2 \, a_{np}~n_n~n_p
~. \label{pB-multi}
\end{eqnarray}%
Here the shifted chemical potentials $\mu_n^*$ and $\mu_p^*$ are given by
\eq{
\mu_n^* & = \mu_B - b_{nn} \, p_{n}^{\rm id}( T,\mu_n^*) - b_{np} \, p_{p}^{\rm id}( T,\mu_p^*) + 2 \, a_{nn} \, n_n + 2 \, a_{np} \, n_p~,\\
\mu_p^* & = \mu_B + \mu_Q - b_{np} \, p_{n}^{\rm id}( T,\mu_n^*) - b_{nn} \, p_{p}^{\rm id}( T,\mu_p^*) + 2 \, a_{np} \, n_n + 2 \, a_{nn} \, n_p~,
}
The generalized equations for neutron and proton densities are~\cite{Vovchenko:2017zpj}
\eq{\label{nB-multi}
n_n(T,\mu_B,\mu_Q)&~=~\frac{n_{n}^{\rm id}(T,\mu_n^*)~\left[1+(b_{nn}-b_{np})~n_{p}^{\rm id}(T,\mu_p^*)\right]}
{1~+~b_{nn}~[n_{n}^{\rm id}(T,\mu_n^*)+n_{p}^{\rm id}(T,\mu_p^*)]~+~(b_{nn}^2 - b_{np}^2)~n_{n}^{\rm id}(T,\mu_n^*)~n_{p}^{\rm id}(T,\mu_p^*)}~,\\\label{nQ-multi}
n_p(T,\mu_B,\mu_Q)&~=~\frac{n_{p}^{\rm id}(T,\mu_p^*)~\left[1+(b_{nn}-b_{np})~n_{n}^{\rm id}(T,\mu_n^*)\right]}
{1~+~b_{nn}~[n_{n}^{\rm id}(T,\mu_n^*)+n_{p}^{\rm id}(T,\mu_p^*)]~+~(b_{nn}^2 - b_{np}^2)~n_{n}^{\rm id}(T,\mu_n^*)~n_{p}^{\rm id}(T,\mu_p^*)}~.
}
If all QvdW parameters are assumed to be equal, $a_{nn}=a_{np}$, $b_{nn}=b_{np}$, Eqs.~(\ref{pB-multi})-(\ref{nQ-multi}) reduce to the model defined by Eqs.~(\ref{pB})-(\ref{nQ}).

The symmetric nuclear matter, $\mu_Q=0$, $n_N=n_Q$, reduces to a single-component system with interaction parameters
\eq{\label{a-b}
a ~=~ \frac{a_{nn} ~+~ a_{np}}{2} \quad \textrm{and} \quad b ~=~ \frac{b_{nn}~ +~ b_{np}}{2}~.
}
The values of $a$ and $b$ are fixed by fitting the known ground state properties of symmetric nuclear matter: $a = 329$~MeV fm$^3$ and $b=3.42$~fm$^3$ (see Sec.~\ref{sec:congruent-A}).
Therefore, the model has two free parameters: $a_{np}/a_{nn}$ and $b_{np}/b_{nn}$. 
We fix these parameters to reproduce the constraints on the symmetry energy and its slope at normal nuclear density.
We take the following values: $a_{np}/a_{nn}=2.5$  and $b_{np}/b_{nn}=1.7$. This yields symmetry energy $J = 32$ MeV and its slope $L = 51$ MeV which are consistent with empirical constraints \cite{Lattimer:2012nd,Kortelainen:2010hv,Tamii:2011pv,Chen:2010qx,Tsang:2008fd,Trippa:2008gr,Steiner:2010fz}.

\begin{figure}[h!]
\includegraphics[width=0.70\textwidth]{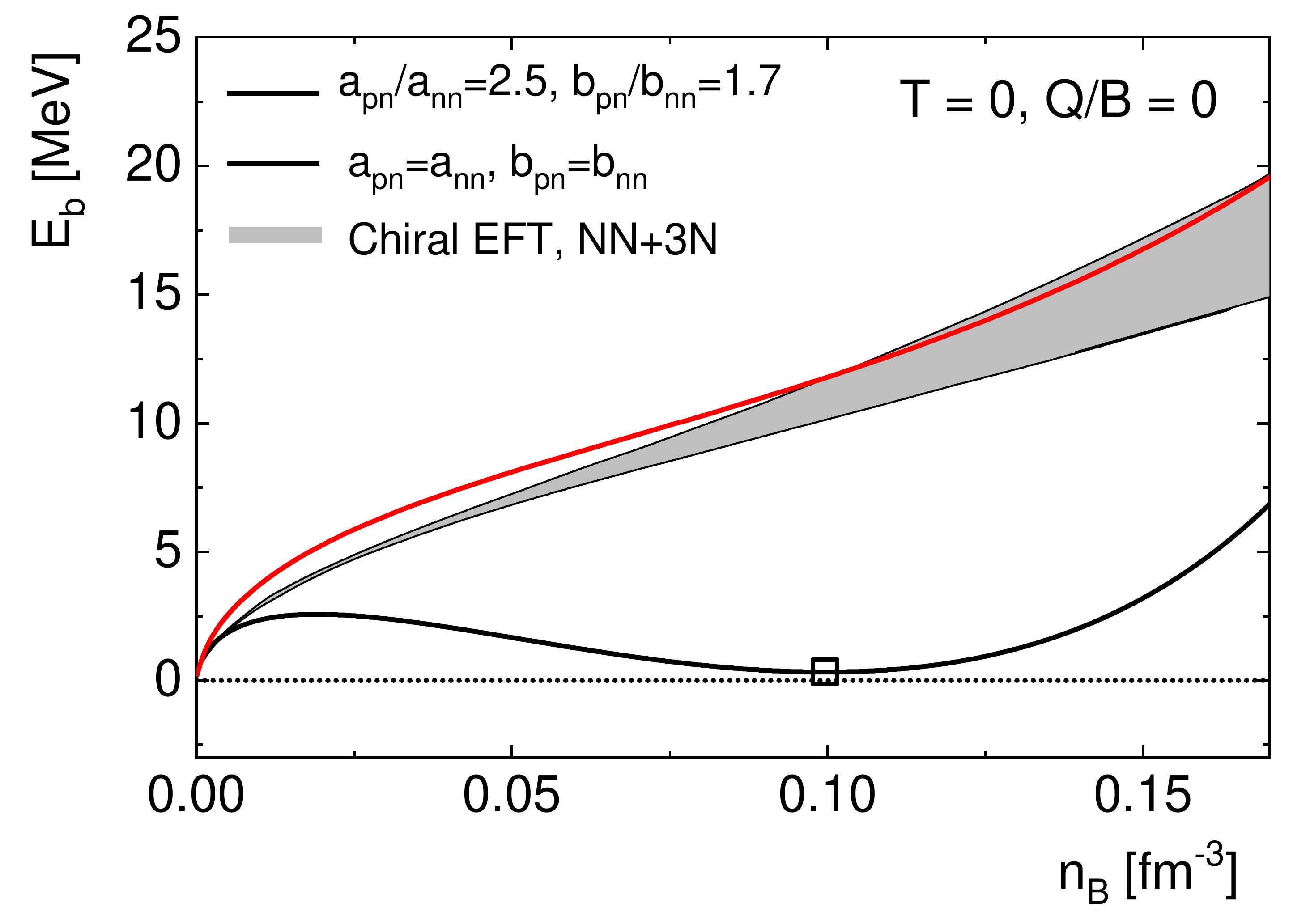}
\caption{\label{eb-n-multi} Dependence of the binding energy on baryon density at zero temperature. 
Black line is the same as in Fig. \ref{eb-n}.
The red line represents the QvdW model with isospin-dependent interaction parameters. The gray band shows result of chiral effective field theory \cite{Hebeler:2013nza}.
}
\end{figure}

The red line in Fig.~\ref{eb-n-multi} shows the dependence of the binding energy on $n_B$ at zero temperature for pure neutron matter within QvdW model with $a_{np}/a_{nn}=2.5$  and $b_{np}/b_{nn}=1.7$. The gray band represents result for pure neutron matter within chiral effective mean field model \cite{Hebeler:2013nza} with NN and 3N interactions and a renormalization-group evolution. The width of the band is mainly due to uncertainties in 3N forces. In the considered case $a_{np}/a_{nn}=2.5$ and $b_{np}/b_{nn}=1.7$ the minimum in the binding energy as a function of density is absent for $Q/B\lesssim 0.08$.

\begin{figure}[h!]
\includegraphics[width=0.49\textwidth]{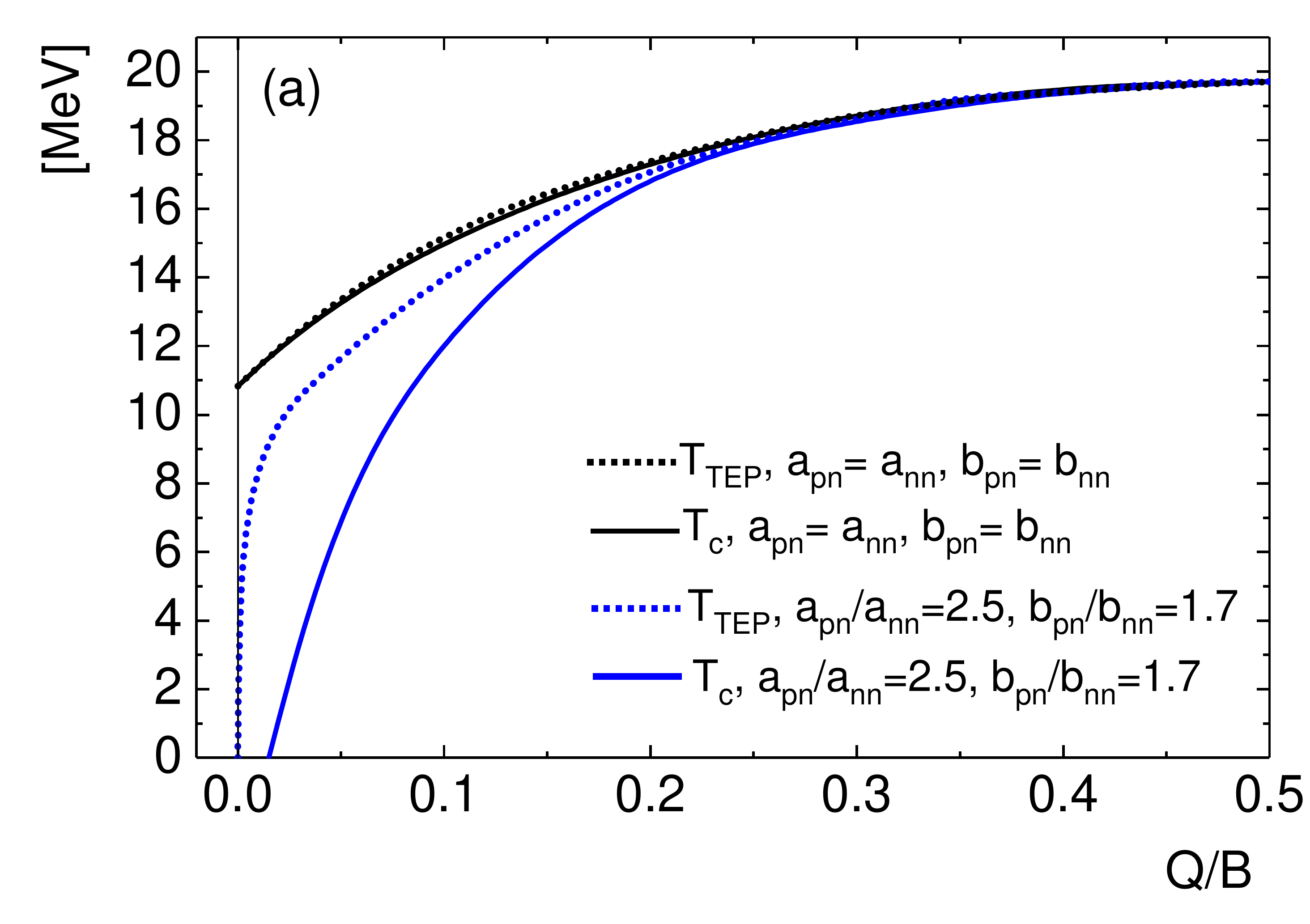}
\includegraphics[width=0.49\textwidth]{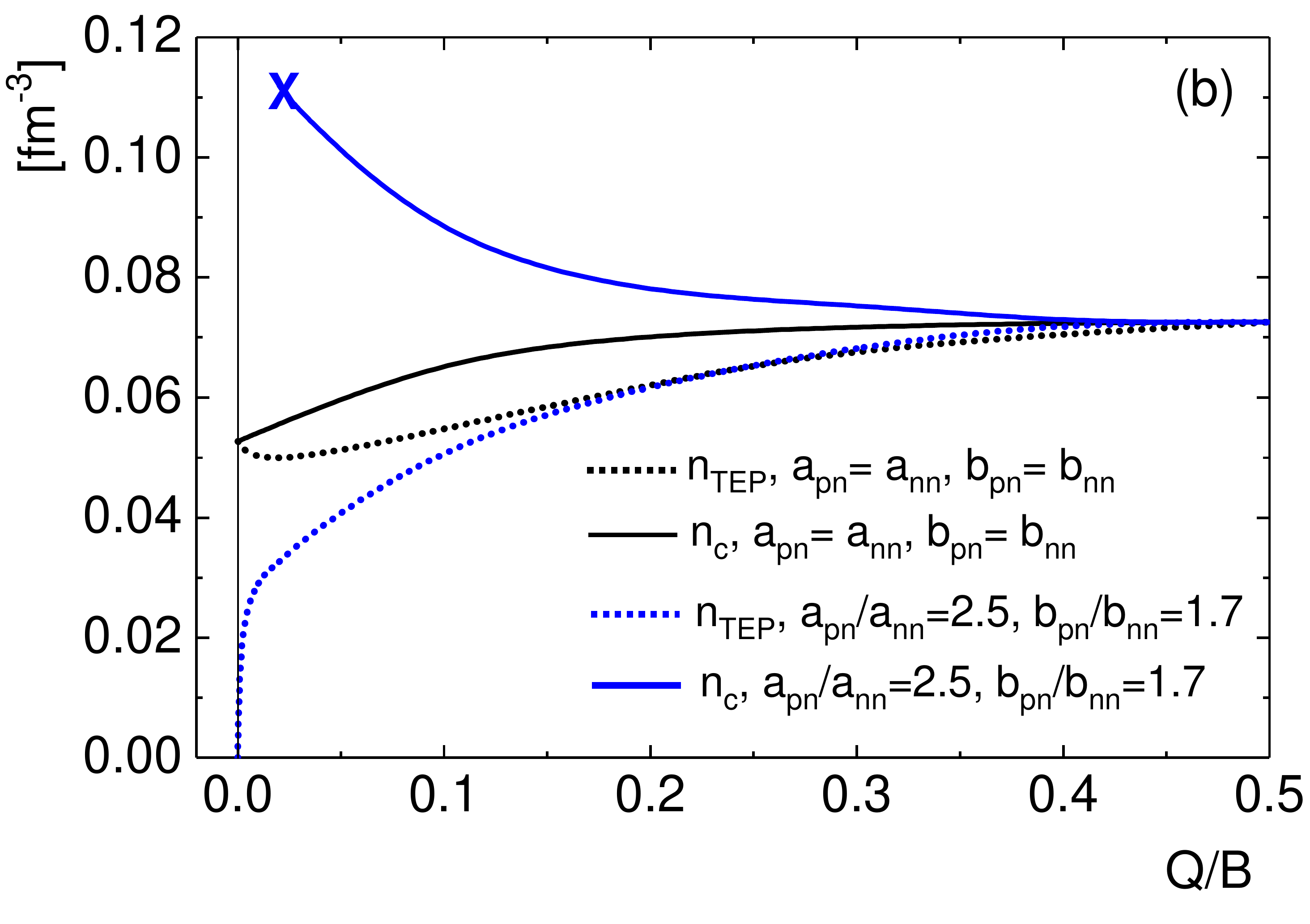}
\caption{\label{fig-Tc} ($a$) $T_c$ and ($b$) $n_c$ as functions of $Q/B$. Black and blue lines correspond, respectively, to the cases $a_{np}=a_{nn}=a$, $b_{np}=b_{nn}=b$ and $a_{np}/a_{nn}=2.5$, $b_{np}/b_{nn}=1.7$. Solid and dotted lines are the lines of, respectively, critical points and temperature endpoints.
}
\end{figure}

Figures \ref{fig-Tc} ($a$) and ($b$) show, respectively, the dependencies of $T_c$ and $n_c$ on $Q/B$, calculated within the QvdW model for $a_{np}=a_{nn}=a$, $b_{np}=b_{nn}=b$~(black lines) and $a_{np}/a_{nn}=2.5$, $b_{np}/b_{nn}=1.7$~(blue lines). 
In the case of symmetric nuclear matter, $Q/B=0.5$, the two parameter sets yield identical results, as follows from the condition~(\ref{a-b}).
For a sufficiently symmetric system, $Q/B\gtrsim 0.3$, the consideration of isospin dependence in interaction parameters does not influence significantly the locations of CP and TEP.

The situation is different in the case of a large asymmetry.
In the case of isospin-independent interactions, the CP is present for all values of $Q/B$, even in the case of a pure neutron matter, $Q/B=0$: $T_c=T_{TEP}\simeq 10.83$.
On the other hand, when isospin-dependent interaction parameters are used, the critical temperature approaches zero at a finite $Q/B \simeq 0.015$, and the CP is no longer present at $Q/B \lesssim 0.015$.
The TEP, however, is present for all finite values of $Q/B$, with the limiting behavior $T_{TEP}\rightarrow 0$ at $Q/B\rightarrow 0$.
These observations suggest that for $Q/B \lesssim 0.015$ the CP is absent, however the LGPT is still present, in the form of a retrograde condensation.
In the extreme case of pure neutron matter, $Q/B = 0$, the phase transition disappears completely.

The comparison between isospin-independent and isospin-dependent parameterizations of the QvdW interactions in Fig.~\ref{fig-Tc} shows that the former is appropriate for possible applications to heavy-ion collisions, which are typically characterized by $Q/B \simeq 0.4$.


\section{summary}
\label{summary}

The non-congruent liquid-gas phase transition in asymmetric nuclear matter with two globally conserved charges is studied within the Quantum van der Waals model. The features of those non-congruent phase transitions, such as the continuous phase transformation, a change in the location of the critical point, the separation of the critical point and of the endpoints, and the retrograde condensation have been analyzed. The magnitudes of these phenomena tend to zero if the composition of the nuclear matter approaches the composition of either the limit of symmetric nuclear matter ($Q/B=0.5$) or pure neutron matter ($Q/B=0$). 
The scaled variances of the baryonic and the electric charges fluctuations are calculated in the presence of the non-congruent phase transition. The fluctuations of both baryonic and electric charges are divergent in the critical point. 

The Quantum van der Waals model with isospin-dependent interaction parameters, constrained to empirical values of the symmetry energy and its slope, has also been considered. It yields results which are similar to the case of isospin-independent interaction parameters for sufficiently symmetric systems, $Q/B\gtrsim 0.3$, which covers possible applications in heavy-ion collisions. 
However there are important differences for small values of the asymmetry parameter. In particular, the ground state and the phase transition  are absent in pure neutron matter in the case of the isospin-dependent QvdW interactions.

~\\
{\bf Acknowledgments:}
The authors thank D. V. Anchishkin, B. I. Lev, A. G. Magner,  M. Gazdzicki, A. Motornenko, K. Taradiy, A. I. Sanzhur, L. M. Satarov, and G. M. Zinovjev for fruitful discussions and useful comments. 
This research was supported by theme grant of department of physics and
astronomy of NAS of Ukraine: ``Dynamics of formation of spatially
non-uniform structures in many-body systems", PK 0118U003535.
H. St. appreciates the support through the Judah M.
Eisenberg Laureatus Chair at Goethe University, and the  Walter
Greiner Gesellschaft, Frankfurt.

\bibliography{asymmetric-nuclear-matter}

\end{document}